\documentclass[twocolumn]{aastex62}
\usepackage{longtable}
\usepackage{xcolor}

\bibpunct{(}{)}{;}{a}{}{,}
\begin{document}
	
	\defcitealias{Bisigello2016}{B16}
	\defcitealias{Bisigello2017}{B17}
	
	\title{Statistical stellar mass corrections for high-z galaxies observed with \emph{JWST} broad-band filters due to template degeneracies}
	\author{L. Bisigello}
	\affiliation{Kapteyn Astronomical Institute, University of Groningen, P.O. Box 800, 9700 AV, Groningen, The Netherlands.}
\affiliation{University of Nottingham, School of Physics \& Astronomy, Nottingham, NG7 2RD, UK}
	
	\author{K. I. Caputi}
	\affiliation{Kapteyn Astronomical Institute, University of Groningen, P.O. Box 800, 9700 AV, Groningen, The Netherlands.}
	\affiliation{Cosmic Dawn Center (DAWN), Niels Bohr Institute, University of Copenhagen, Juliane Maries vej 30, DK-2100 Copenhagen, Denmark}
	
	\author{L. Colina}
	\affiliation{Centro de Astrobiolog\'{\i}a, Departamento de Astrof\'{\i}sica, CSIC-INTA, Cra. de Ajalvir km.4, 28850 - Torrej\'on de Ardoz, Madrid, Spain}
	\affiliation{Cosmic Dawn Center (DAWN), Niels Bohr Institute, University of Copenhagen, Juliane Maries vej 30, DK-2100 Copenhagen, Denmark}
	
	\author{P. G. P\'erez-Gonz\'alez}
	\affiliation{Departamento de Astrof\'{\i}sica, Facultad de CC. F\'{\i}sicas, Universidad Complutense de Madrid, E-28040 Madrid, Spain; Associate Astronomer at Steward Observatory, The University of Arizona}
	\affiliation{Centro de Astrobiolog\'{\i}a, Departamento de Astrof\'{\i}sica, CSIC-INTA, Cra. de Ajalvir km.4, 28850 - Torrej\'on de Ardoz, Madrid, Spain}
	
	\author{ A. Koekemoer}
	\affiliation{Space Telescope Science Institute, 3700 San Martin Drive, Baltimore, MD 21218, USA}
	
	\author{O. Le F\`evre}
	\affiliation{Aix Marseille Universit\'e, CNRS, LAM (Laboratoire d'Astrophysique de Marseille), UMR 7326, 13388, Marseille, France}
	
	\author{N. Grogin}
	\affiliation{Space Telescope Science Institute, 3700 San Martin Drive, Baltimore, MD 21218, USA}
	
	\author{H. U. N\o rgaard-Nielsen}
	\affiliation{National Space Institute (DTU Space), Technical University of Denmark, Elektrovej, DK-2800 Kgs. Lyngby, Denmark}
	
	\author{P. van der Werf}
	\affiliation{Sterrewacht Leiden, Leiden University, PO Box 9513, 2300 RA, Leiden, The Netherlands}

	\shorttitle{Bisigello et al.: }
	
	\shortauthors{Bisigello et al.}
	
	\begin{abstract}
		Stellar masses in future \textit{James Webb Space Telescope} (\textit{JWST}) deep blank-fields will be mainly derived fitting the spectral energy distribution with theoretical galaxy templates. 
		We investigate the uncertainties and biases of the stellar masses derived by using the \textit{LePhare} code for spectral energy distribution fitting and the \textit{Yggdrasil} theoretical templates. We consider a sample of mock galaxies at z$=$7-10 with mock JWST observations with S/N$_{F150W}\geqslant$ 10.  Our goal is to provide a list of statistical stellar mass corrections to include on the stellar mass derivation for different output galaxy properties and JWST filter combinations to correct for template degeneracies. Median statistical stellar mass corrections vary from -0.83 dex to 0.87 dex, while 25$\%$ (75$\%$) quartiles range from -0.83 dex (-0.67 dex) to 0.51 dex (0.88 dex), depending on filter combinations and galaxy models. The most challenging cases are galaxies with nebular emission lines, especially the ones that are wrongly identified as galaxies without, relative dust-free galaxies and galaxies with small metallicities (i.e. Z$=1/50Z_{\odot}$). The stellar mass estimation of galaxies correctly identified without emission lines is generally fine, except at z$=$10 when considering only the 8 NIRCam bands, which make the MIRI bands very valuable. We have tested our stellar mass corrections using the public JAGUAR galaxy catalogue, deriving that the average discrepancy in the recovered stellar mass distribution decreases by 20-50$\%$ at z$>$7 after the correction. We found that without the stellar-mass corrections the number of low-mass galaxies (M$^{*}<10^{7}M_{\odot}$) is overestimated, which can potentially lead to systematic errors in the calculation of the galaxy stellar mass function faint-end slope at high z.

	\end{abstract}
	
	\keywords{galaxies: high-redshift; galaxies: photometry; galaxies: stellar mass} 
	
	\section{Introduction} \label{sec:intro}
	The stellar mass is one of the most fundamental property of galaxies, as it has a central role in galaxy evolution \citep{Peng2010}. Its importance is evident from the numerous relations that are present between galaxy stellar mass and other galaxy properties, such as star formation rate \citep[e.g.][]{Brinchmann2004,Noeske2007,Rodighiero2011,Whitaker2014,Tasca2015,Bisigello2018,Boogaard2018} and metallicity \citep[e.g. ][]{Tremonti2004,Erb2006,Maiolino2008,Maier2015}. \par
	
	By fitting the spectral energy distribution (SED) of galaxies derived from broad-band photometry with theoretical or empirical templates, it is possible to derive a broad set of galaxy properties, among which there is also the stellar mass. This is a powerful technique, because it allows for deriving properties of galaxies up to high redshift, thanks to the possibility to obtain broad-band photometry for large samples of faint galaxies. Comparatively, with spectroscopy it is possible to have a more reliable estimate of some of these galaxy properties, but spectroscopic observations require more integration time than photometry and are usually possible only for relatively bright galaxies. \par
	
	It has been shown in previous works that, once the redshift of a galaxy is well known, the stellar mass is one of the most robust parameters derived from the SED fitting \citep[e.g.][]{Caputi2015}. However, precise stellar masses are particularly difficult to estimate in some situations. 
	Indeed, galaxies on-going strong episodes of star-formation are dominated by a young stellar population and the stellar mass derivation is strongly affected by uncertainties in the age estimation. In addition, these young and star-forming galaxies may have numerous nebular emission lines with high equivalent widths that contaminate the broad-band observations and affect the derived stellar mass \citep[e.g.][]{Stark2013,Santini2015,Caputi2017}. These nebular emission lines are generally visible as a flux excess in some bands, but in some extreme cases, they are so numerous to boost the flux of contiguous bands and mimic a higher continuum. This results in a degeneracy between young galaxies with numerous emission lines and more massive, older galaxies \citep{Bisigello2017}. \par
	In addition, observations at rest-frame near-ultraviolet (UV) are affected by dust-extinction and the mass-to-light ratio at these wavelengths is sensitive to small differences in the stellar population age. Therefore, it is necessary to have observations at wavelengths longer than $4000\,\AA$ to have a good estimate of the stellar mass. This means that at very high-z it is necessary to observe at near-infrared wavelengths to have a good stellar mass estimation and this will be indeed possible in the next future with the \textit{James Webb Space Telescope} (\emph{JWST}\footnote{http://www.jwst.nasa.gov},\citealt{Gardner2009}).\par
	\emph{JWST} is among the most promising facilities of the next years. It has a 6.5-meter primary mirror and four instruments on board observing at near- and mid-IR wavelengths (0.6-28 $\mu$m) with imaging, spectroscopic and coronographic modes. In particular, the Near Infrared Camera \citep[NIRCam;][]{Rieke2005} is an imaging camera covering from 0.6 to 5 $\mu$m with different broad, intermediate and narrow-band filters. On the other hand, the complementary Mid Infrared Instrument \citep[MIRI;][]{Rieke2015,Wright2015} has nine broad-band filters that cover between 5 and 28 $\mu$m. These two imaging cameras will be the main instruments used to carry out deep blank-field imaging surveys with JWST to detect high-z galaxies. \par
	
	The properties of galaxies observed in these deep blank-field imaging surveys will be mainly derived using SED fitting. In \citealt{Bisigello2016,Bisigello2017} (hereafter B16 and B17, respectively) we created and analyzed a sample of mock galaxies at z$=$7-10 to study how different galaxy properties will be derived using SED fitting and different NIRCam and MIRI broad-band filter combinations. In particular, in \citetalias{Bisigello2017}, we show that stellar masses may be particularly difficult to estimate for specific filter combination and particular galaxy templates, i.e. galaxies with numerous nebular emission lines and galaxies at z$=$10, in particular if mid-IR observations are not available. The aim of this paper is to derive and analyze the stellar mass offset for each specific galaxy template and for different JWST broad-band filter combination, to compensate for template degeneracies or lack of wavelength coverage. These corrections will be extremely useful to statistically correct the stellar mass of large sample of high-z galaxies that will be observed using different combination of \textit{JWST} broad-band filters in the next future to study, for example, the stellar mass function at z$>$7.\par
	
	The paper is structured as follows. In section \ref{sec:sample} we describe the analyzed sample of mock galaxies, the photometry extraction in the pertinent NIRCam and MIRI bands and the stellar mass derivation. We present our stellar mass corrections for different JWST broad-band filter combinations in section \ref{sec:results}. In addition, we analyze the stellar mass correction respect to other galaxy properties, such as redshift, age, color excess, metallicity and star formation history (SFH). In section \ref{sec:application} we give practical information on how to include the derived stellar mass correction on future studies and in section \ref{sec:comparison} we apply and test the derived corrections to a galaxy sample. Finally, in section \ref{sec:conclusions} we summarize our main findings and conclusions. Throughout this paper, we consider a cosmology with $H_0=$70~$\rm km \, s^{-1} \, Mpc^{-1}$, $\Omega_M=0.27$, $\Omega_\Lambda=0.73$ and a \cite{Kroupa2002} initial mass function (IMF). All magnitudes refer to the AB system \citep{Oke1983}. \par

	\section{Sample} \label{sec:sample}
	\subsection{Sample selection}
	Our study is based on a sample of 750 simulated galaxies at z$=$7-10, presented in \citetalias{Bisigello2016} and derived from the \textit{Yggdrasill} population synthesis code \citep{Zackrisson2011}, which is specifically built to describe high-z galaxies. These templates have solar and sub-solar metallicities, step function star formation histories, color excess between 0 and 0.25 mag considering a \cite{Calzetti2000} reddening law. We apply the same dust attenuation for continuum and nebular emission lines, however,considering different dust extinction values for emission lines and continuum may vary some of the template degeneracies but does not create any systematic shift in the statistical stellar mass correction. Ages are between 0.01 and 0.6 Gyr and consistent with the age of the Universe at z$=$7 to 10 and we consider a \cite{Kroupa2002} IMF. Nebular continuum and emission lines are already incorporated in the \textit{Ygdrasill} templates when the galaxy is star-forming and the covering factor is not zero. We consider two different covering factors, corresponding to galaxies without nebular emission lines, f$_{cov}=$0, and with the maximum contribution from the nebular lines, f$_{cov}=$1. These values correspond to a Lyman continuum escape fraction of 1 and 0, respectively, as f$_{esc}=1-$f$_{cov}$. \par
	
	\subsection{Mock JWST photometry}
	For all 750 templates we have mock observations for the 8 NIRCam broad bands and the two MIRI broad-bands F560W and F770W, as explained in detailed in \citetalias{Bisigello2016}. These mock observations are obtained convolving each template with the corresponding \textit{JWST} filter. All simulated galaxies are normalized at 29 AB mag at 1.5 $\mu$m, which corresponds to the pivot wavelength of the F150W NIRCam band. The signal-to-noise values of 10 and 20 are considered for the F150W band and the same integration time is assumed for all the other NIRCam bands. For the MIRI bands, we consider the same signal-to-noise of the F150W NIRCam band, but for a magnitude brighter, i.e. 28 AB mag. This has been done to take into account different sensitivities between the NIRCam and MIRI bands, due to different detector technologies. We consider only a S/N$\geq$10, because for more than 99$\%$ of the simulated galaxies at z$=$7-10 with signal-to-noise of 10 the photometric redshift is well recovered, i.e. $|z_{phot}-z_{fiducial}|/(1 + z_{fiducial} )\leqslant$0.15, already using only 8 NIRCam bands. Therefore, errors in the stellar masses are only due to degeneracies between templates and not to a drastically incorrect redshift estimation.	For each simulated galaxy, we have 100 mock observations in the considered NIRCam and MIRI bands, derived randomizing each flux inside the error bars, for a total sample of 75000 mock observations. The results presented in this paper are valid also for galaxies with stellar masses different from the analyzed ones, i.e. corresponding to 29 AB mag at 1.5 $\mu$m, as long as the S/N$\geqslant$10 and the SED shapes are the same. In general, it is necessary to consider with caution the application of these results on stellar masses derived by using templates with extremely different prescriptions for nebular emission lines, star-formation histories or the general SED shape. In addition, it is necessary to take into account that the stellar mass offset analysed in this work are due only to template degeneracies and are therefore non-exhaustive as, for example, additional errors are expected due to differences between idealize theoretical templates and real galaxy SEDs. \par
	
	\subsection{Galaxy properties derivation}\label{sec:PropDerivation}
	The redshift and stellar mass recovery for these simulated galaxies have been derived and analyzed in \citetalias{Bisigello2016} and \citetalias{Bisigello2017}, respectively. In particular, galaxy properties have been obtained for different combination of \textit{JWST} broad-band observations:
	\begin{itemize}
		\item 8 NIRCam broad bands
		\item 8 NIRCam broad bands and 2 MIRI bands (F560W and F770W)
		\item 8 NIRCam broad bands and MIRI F560W only
		\item 8 NIRCam broad bands and MIRI F770W only
	\end{itemize}
	For all these different filter combinations, we derive the stellar mass and the photometric redshift using the public code \textit{LePhare} \citep{Arnouts1999,Ilbert2006} and considering a large set of possible output templates, including the ones used to derive the mock observations. In particular, we consider a wide range of color excess, i.e. from 0 to 1 mag with a step 0.05 mag, redshifts from 0 to 11, ages from 0.01 Gyr up to 5 Gyr and consistent with age of the Universe. We also include \citet{B&C2003} templates with exponentially declining star formation histories with the same values for redshift, age and color excess of the \textit{Yggdrasill} templates and with emission lines. However, almost all galaxies are best fitted by \textit{Yggdrasill} and the very few exceptions result on extremely large redshift errors, i.e. output z$<$2. We include two covering factors, 0 and 1, but only when the star-formation is on-going, as this parameter does not influence the SED when the galaxy is not star-forming. For more details on the set of template parameters used to derive the stellar mass and the procedure to estimate it we refer to \citetalias{Bisigello2017}. \par
	
	In this paper, we present the stellar mass correction for galaxies corresponding to each combination of output properties. We limit our analyses only to mock galaxies with a combination of output parameters that are present among the input ones, but that are not necessarily the correct one.The stellar mass correction for galaxies with output parameter values not present among the input ones would be overestimated, because the output galaxy model is always different from the input one by construction and, therefore, the stellar mass is never correct. We divide color excess and redshifts in bins of $\Delta (E(B-V))=$0.1 mag and $\Delta z=$1 centered around the values present among the mock observations.\par
	Overall we remove from the sample galaxies with output redshifts below 6.5 and above 10.5, which correspond to $<$1.2$\%$ and $<$2.3$\%$ with all band combinations. Moreover, we do not consider galaxies with output color excess larger that 0.3 mag, that correspond to 1.3$\%$ of the sample when only NIRCam broad-bands are considered and $<$0.5$\%$ when at least one of the two MIRI band is included. The full set of galaxy properties considered in this work is listed in Table \ref{tab:param} and their input distributions are shown in figure \ref{fig:input_parm}. Given these points, the final samples consist of 71257 (95$\%$ of the original catalogue) galaxies with NIRCam observations, 72897 (97$\%$ of the original catalogue) galaxies with NIRCam and F560W band observations, 72295 (96$\%$ of the original catalogue) galaxies with NIRCam and F770W band observations, and 73741 (98$\%$ of the original catalogue) galaxies with observations in the 8 NIRCam and 2 MIRI bands. \par
	
	The stellar mass of each mock galaxy is obtained by \textit{LePhare} by scaling the template considering all bands, while the input templates are scaled to match a magnitude 29 AB at 1.5 $\mu$m. This difference may result on a bias on the derived stellar mass. Therefore, we analyze the recovered stellar mass for a subsample of mock galaxies for which all other galaxy parameters are perfectly recovered, i.e. the input template is correctly recognized. We find that a small bias is present and stellar masses are, on average, overestimated of $\sim$5-6$\%$. 
	Hereafter, all the output stellar masses are corrected by this general bias of -0.025 dex, to remove the dependence on the used SED fitting code. \par 
	
	\subsection{Dependence of the stellar mass corrections on the chosen input parameters}
\begin{figure*}[ht!]
		\center{
			\includegraphics[trim={0cm 0cm 0cm 0cm},clip,width=0.9\linewidth, keepaspectratio]{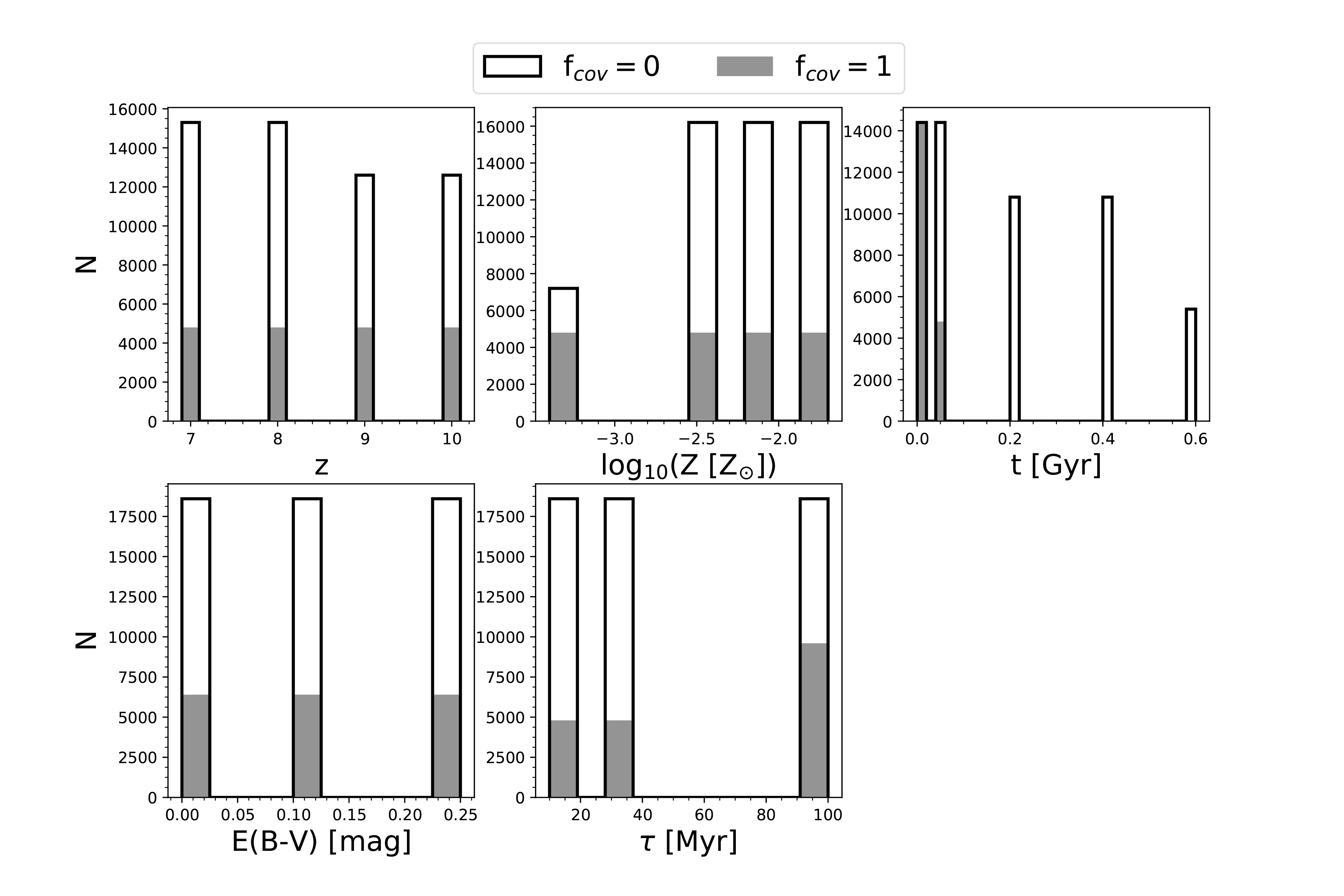}}
		\caption{Distribution of each input galaxy parameter used to derive the galaxy sample considered in this work. \textit{From upper left to bottom right:} redshift, metallicity, galaxy age (time from the beginning of star formation), colour excess, duration of the star formation episodes. Each parameter distribution is shown separately for galaxies with f$_{cov}=$0 (\textit{empty histograms}) or f$_{cov}=$1 (\textit{filled histograms}) \label{fig:input_parm}}
	\end{figure*}

	The stellar mass corrections derived here depend on the assumed input galaxy population, which is derived assuming the parameters listed in Table \ref{tab:param} and following the distributions shown in Fig. \ref{fig:input_parm}. All parameter combinations have been included, except for templates with ages longer than the Universe age at the considered redshift and the lowest metallicity (0.02 Z$_{\odot}$) which is considered only for templates with ages shorter than 0.2 Gyr. The results presented in this paper are affected by the used parameters in two ways. First, templates described by different parameters may arise additional degeneracies that are not taken into account here, therefore the stellar mass corrections presented needs to be considered as not exhaustive. Second, if the input distribution of each parameter is extremely different, the overall degeneracies will remain the same, but each probability may be different. \par 
	Under the assumption that nebular emission lines depend on metallicity, dust extinction, covering factor, age and SFH, but not on stellar mass, the stellar mass corrections derived in this paper do not depend on the assumed stellar mass distribution. Indeed, if two templates well represent some observations, they would represent equally well the same observations scaled by an arbitrary factor, as long as the S/N is the same and the scaling factor is the same for all observations. The absolute stellar mass would change because of the re-scaling, but the stellar mass correction, which is a relative quantity, would remain the same. For this reason our results do not depend on any assumption on the input stellar mass distribution and can be generally applied to any statistical sample. However, we advise against using the corrections presented here for samples biased on any of the other input parameters, i.e. a sample containing only emission-line galaxies. In case of a biased sample, the corrections presented here can be anyway considered to identify possible biases affecting the stellar mass.\par  
	
	\begin{deluxetable}{cc}
		\tablecaption{Output values of different galaxy properties considered in this work. \label{tab:param}} 
		\tablecolumns{3}
		\tablenum{1}
		\tablewidth{0pt}
		\tablehead{
			\colhead{Parameter} &
			\colhead{Values }
		}
		\startdata
		metallicity &  Z$_{\odot}$,0.4Z$_{\odot}$,0.2Z$_{\odot}$,0.02Z$_{\odot}$\tablenotemark{a}\\
		SFH type &  step function \\
		SFH [Gyr] & 0.01,0.03,0.1\\
		$f_{\rm cov}$ & 0,1\tablenotemark{b}\\
		E(B-V)\tablenotemark{c} &  0,0.05,0.1,0.15,0.2,0.25,0.3 \tablenotemark{d} \\
		age [Gyr] &  0.01,0.05,0.2,0.4,0.6\tablenotemark{e}\\
		z &  6.5,6.55,6.70,...,10.40,10.45,10.5 \tablenotemark{f}\\
		\enddata
		\tablenotetext{a}{for this metallicity we consider only ages $t<$0.2~Gyr.}
		\tablenotetext{b}{templates of old galaxies with no ongoing star formation do not change with the covering factor, so, for these galaxies, we consider only f$_{cov}=$0.}
		\tablenotetext{c}{following Calzetti et al. reddening law \citep{Calzetti2000}}
		\tablenotetext{d}{Results are showned in bins of $\Delta (E(B-V))=$0.1 mag.}
		\tablenotetext{e}{we consider this age only up to redshift $z=8$.}
		\tablenotetext{f}{Results are showned in bins of $\Delta z=1$.}
	\end{deluxetable}
	
	\section{Results} \label{sec:results}
	In this section we present the statistical stellar mass corrections for the different \textit{JWST} broad-band filter combinations. These can be used to statistically correct the stellar mass estimate for degeneracies arising from the SED fitting procedure. Statistical stellar mass corrections are derived comparing the output stellar mass with the input one, i.e. $log_{10}(M^{*}_{out})-log_{10}(M^{*}_{in})$. The median statistical stellar mass corrections are in general small, as shown in Fig. \ref{fig:DM-z}-\ref{fig:DM-Z}, but the quartile ranges are large, therefore some galaxy models may need significant stellar mass corrections. In particular, 25$\%$ quartiles range from -0.83 dex to 0.51 dex, while the 75$\%$ quartiles are between -0.51 dex to 0.88 dex. \par
	In the next sections, we analyze how these stellar mass corrections are related to other output parameters, i.e. covering factor, redshift, age, color excess, star formation history and metallicity, to understand for which SED templates the stellar mass estimation is particularly challenging. All plots refers to mock galaxies with S/N$_{F150W}=$10, but results are similar for S/N$_{F150W}=$20. We include among the online material the detail stellar mass offset distribution of both signal-to-noise values.
	
	\subsection{Variation of statistical stellar mass correction with covering factor}
	We separate between galaxies with output f$_{cov}=1$ or f$_{cov}=0$, i.e. star-forming galaxies with emission lines or galaxies without emission lines. Among all galaxies with output f$_{cov}=1$, 22$\%$ have input f$_{cov}=0$ when considering only NIRCam bands. Among these galaxies, 83$\%$ (41\%) have stellar mass offsets larger than 0.1 dex (0.3 dex) in modulus, with stellar masses that tend to be underestimated, even up to 10 times. Here, we do not investigate the equivalent width of the nebular emission lines, therefore galaxies there are wrongly identified with f$_{cov}=1$ may have very low output value of equivalent widths. On the other hand, even the population of galaxies correctly identified as galaxies with f$_{cov}=1$ shows large stellar mass offsets, but with less frequency, i.e. 40$\%$ (22$\%$) with stellar mass offsets larger than 0.1 dex (0.3 dex) in modulus. 
	The inclusion of the MIRI bands decreases the percentage of galaxies with input f$_{cov}=0$ to 8$\%$ of all galaxies with output f$_{cov}=1$. \par 
	Among galaxies with output f$_{cov}=0$, $\sim5\%$ have input f$_{cov}=1$, when considering only NIRCam bands. This percentage only slightly changes to 3$\%$ when adding the MIRI broad bands. Independently by the considered filter combination, the stellar mass for the majority of these galaxies is overestimated, even up to 10 times in some cases, and more than 90$\%$ (60$\%$) of them have stellar mass offsets larger than 0.1 dex (0.3 dex) in modulus. This is extremely high compared with the fraction of galaxies correctly identified as galaxies without emission lines, for which only 12$\%$ (2$\%$) have stellar mass errors larger than 0.1 dex (0.3 dex) in modulus, already considering only the 8 NIRCam bands. \par
	Overall, it is evident that the big offsets on the stellar mass arise from galaxies for which the covering factor, i.e. the presence of nebular emission lines, is wrongly recognized. However, even galaxies correctly identified as galaxies with nebular emission lines may also have large stellar mass statistical corrections. The inclusion of the MIRI bands generally reduces the fraction of galaxies wrongly identified as galaxies with nebular emission lines and, therefore, it reduces the fraction of galaxies with extreme stellar mass offsets.
	
	\subsection{Variation of statistical stellar mass correction with redshift}
\begin{figure*}[ht!]
		\center{
			\includegraphics[trim={0cm 0cm 0cm 0cm},clip,width=0.9\linewidth, keepaspectratio]{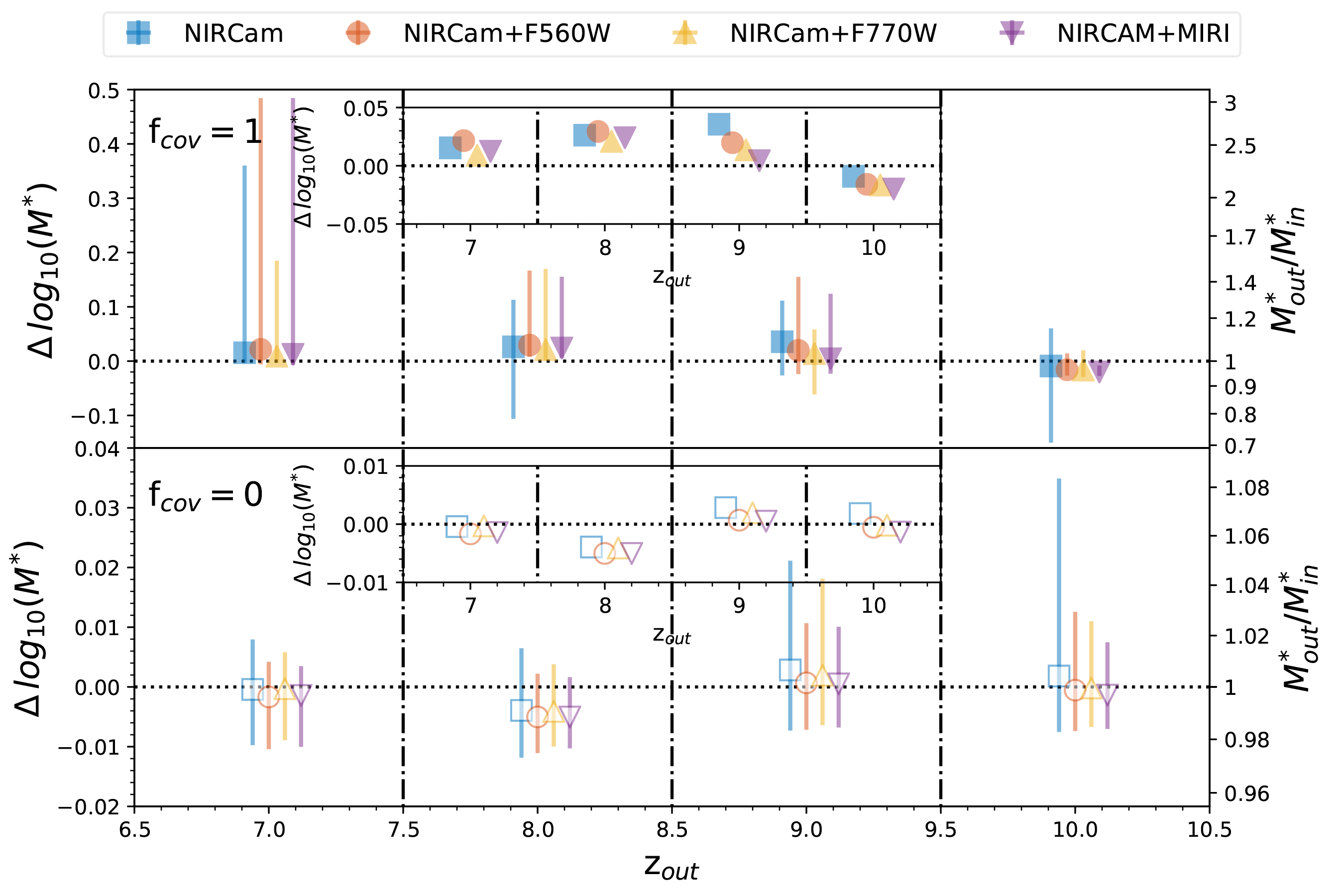}}
		\caption{Median statistical stellar mass correction for different output redshift bins and broad-band filter combinations: 8 NIRCam broad bands (\textit{blue squares}); 8 NIRCam broad bands and MIRI F560W only (\textit{red circles}); 8 NIRCam broad bands and MIRI F770W only (\textit{yellow triangles});  8 NIRCam broad bands, MIRI F560W and MIRI F770W (\textit{purple upside-down triangles}). Galaxies are divided depending on their covering factor: f$_{cov}=$1 (\textit{top}) and f$_{cov}=$0 (\textit{bottom}). The vertical dotted lines indicate the edges of considered the redshift bins. Error bars correspond to the 25$\%$ and 75$\%$ quartiles. Points are slightly offset horizontally respect from each other for illustrative purposes. The inset plots show a zoom-in on the median absolute values \label{fig:DM-z}}
	\end{figure*}

	In Figure \ref{fig:DM-z} we show the median statistical stellar mass correction for different redshift bins and filter combinations. We highlight once again, that here we analyze output redshift values that do not necessarily correspond to the input ones, as it will happen in real observations. All redshift bins are centered around the four input redshifts, z$=$7, 8, 9 and 10. We analyze separately galaxies with different covering factors, i.e. f$_{cov}=$0 or 1. \par
	
	Galaxies with output f$_{cov}=$1 are recognized as star-forming galaxies with the maximum contribution from nebular emission lines. They have generally small median stellar mass offset between 0.036 dex and -0.020 dex, with negative values present at z$=$10. However, the distribution of the stellar mass correction is very broad in some case, in particular at z$=$7 where the 75$\%$ percentile is between 0.18 and 0.48 mag, depending on the considered filter. At z$=$7, extreme stellar mass offset are due to a overestimation of the age for the youngest galaxy template. Excluding z$=$7 templates, the addition of at least one MIRI band reduces the median, that is however already small using only the 8 NIRcam broad-bands, or the dispersion of the stellar mass offset, especially at z$=$10. \par
	
	Galaxies with output f$_{cov}=$0 are recognized as galaxies without emission lines, that could be both star-forming and quenched galaxies. The median stellar mass offsets are small, ranging between 0.005 dex and 0.003 dex. The 25$\%$ and 75$\%$ quartiles are always within $\pm$0.04 dex, i.e. stellar mass offsets below 10$\%$, with a light dependence with redshift. The larger distribution is present at z$=$10, considering only NIRCam bands and this is due to the fact that no NIRCam bands purely cover $\lambda>4000\,\AA$ break.\par
	
	Overall, as mentioned also in \citetalias{Bisigello2017}, stellar masses are, on average, difficult to estimate for galaxies with nebular emission lines at all redshifts. Their stellar masses tend to be overestimated due to an overestimation of the galaxy age. In addition, stellar masses are challenging to estimate also for galaxies without nebular emission lines at z$=$10 when considering only NIRCam bands.  \par

	\subsection{Variation of the statistical stellar mass correction with galaxy age}
\begin{figure*}[ht!]
		\center{
			\includegraphics[trim={0cm 0cm 0cm 0cm},clip,width=0.85\linewidth, keepaspectratio]{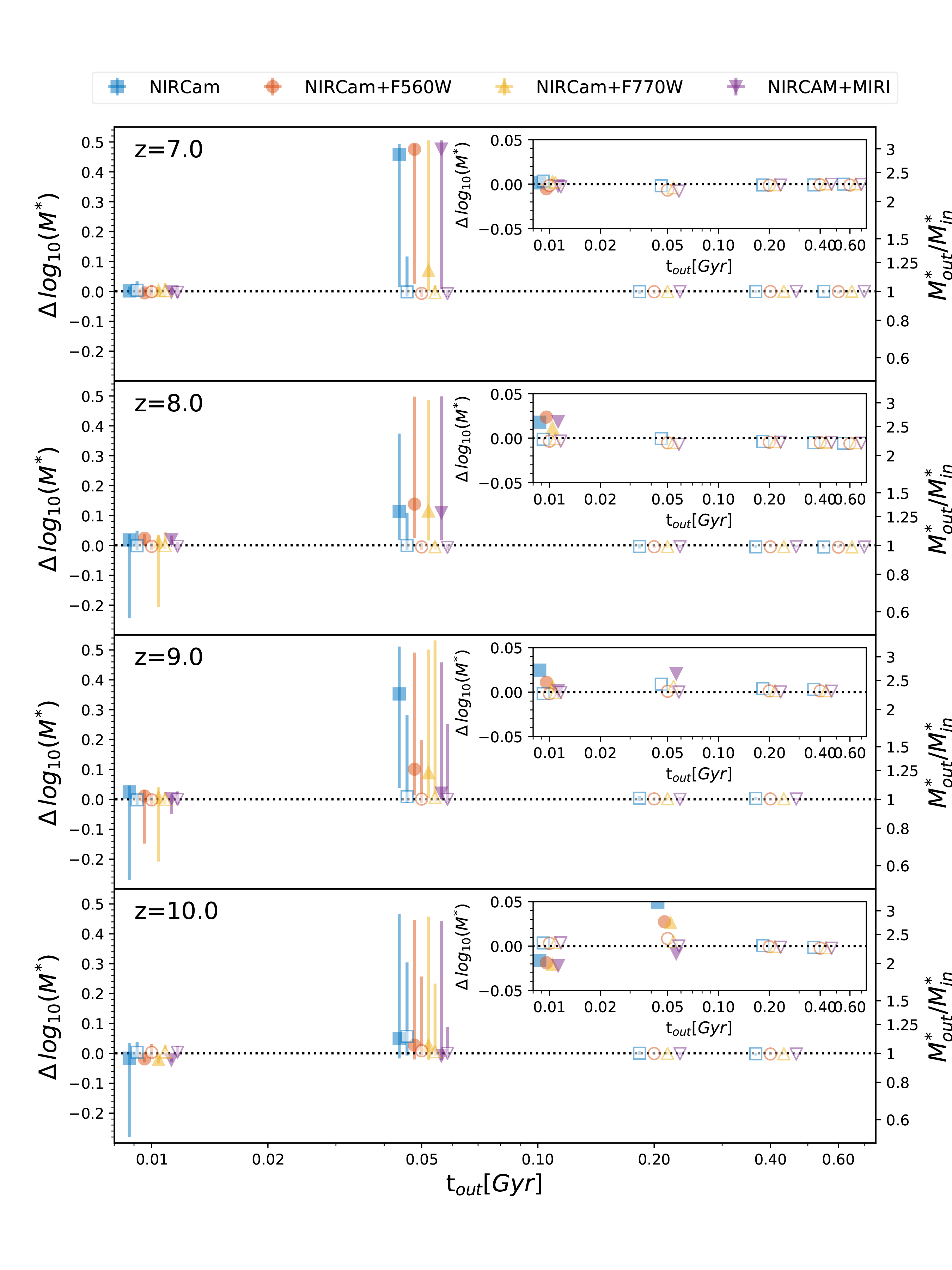}}
		\caption{Median statistical stellar mass correction for different output ages and different output redshift bins. \textit{From top to bottom}: redshifts z = 7, 8, 9 and 10. Different symbols correspond to different broad-band filter combinations: 8 NIRCam broad bands (\textit{blue squares}); 8 NIRCam broad bands and MIRI F560W only (\textit{red circles}); 8 NIRCam broad bands and MIRI F770W only (\textit{yellow triangles});  8 NIRCam broad bands, MIRI F560W and MIRI F770W (\textit{purple upside-down triangles}). Filled symbols indicate galaxies with nebular emission lines, i.e. star forming and with f$_{cov}=$1, while empty symbols indicate galaxies without nebular emission lines, i.e. quiescent galaxies or star-forming galaxies with f$_{cov}=$0. Error bars correspond to the 25$\%$ and 75$\%$ quartiles. Points are slightly offset horizontally respect from each other for illustrative purposes. The inset plots show a zoom-in on the median values. \label{fig:DM-age}}
	\end{figure*}

	Figure \ref{fig:DM-age} shows the median statistical stellar mass correction for different output ages, redshifts and filter combinations. Galaxies with t$_{out}\geqslant$0.20 Gyr have stellar mass corrections within 0.01 dex, i.e. $\sim$2$\%$ error in the stellar mass estimation, including the 25$\%$ and 75$\%$ quartiles and considering all redshifts and JWST broad-band filter combinations. These templates correspond to quiescent galaxies, for which stellar masses are generally correctly estimate. There are no star-forming galaxies with t$_{out}\geqslant$0.20 Gyr, because stars can form only until 0.1 Gyr in the galaxy models included in this work. \par
	On the other hand, galaxies with ages equal or smaller than 0.05 Gyr are star-forming galaxies with or without emission lines, depending on the covering factors, or quiescent galaxies if the template corresponds to a very short period of star-formation. In particular, galaxies with nebular emission lines, i.e. star forming galaxies with f$_{cov}=$1, have stellar masses that are overestimated at t$=$0.05 Gyr, up to 0.5 dex, and generally underestimated at t$=$0.01 Gyr. At z$=$7, a large difference is present on the median offset values derived considering the MIRI/F770W band and both the F560W and F770W MIRI bands, for galaxies with emission lines and t$=$0.05 Gyr. However, the distribution of the statistical stellar mass corrections of these galaxies is bimodal, therefore a small difference in the overall distribution is enough to shift the median of $\sim$0.4 dex.\par
	Galaxies with f$_{cov}=$0, have instead good stellar mass estimation for t$=$0.01 Gyr, but may have overestimated stellar masses for t$=$0.05 Gyr at z$>$8. However, for a small fraction of these galaxies that have output fcov=0 but have nebular emission lines in input, the stellar mass is highly overestimated even up to 1 dex. This generally results from an underestimation of the duration of the star-formation and an overestimation of the age. \par
	
	\subsection{Variation of the statistical stellar mass correction with color excess}
\begin{figure*}[ht!]
		\center{
			\includegraphics[trim={0cm 0cm 0cm 0cm},clip,width=0.8\linewidth, keepaspectratio]{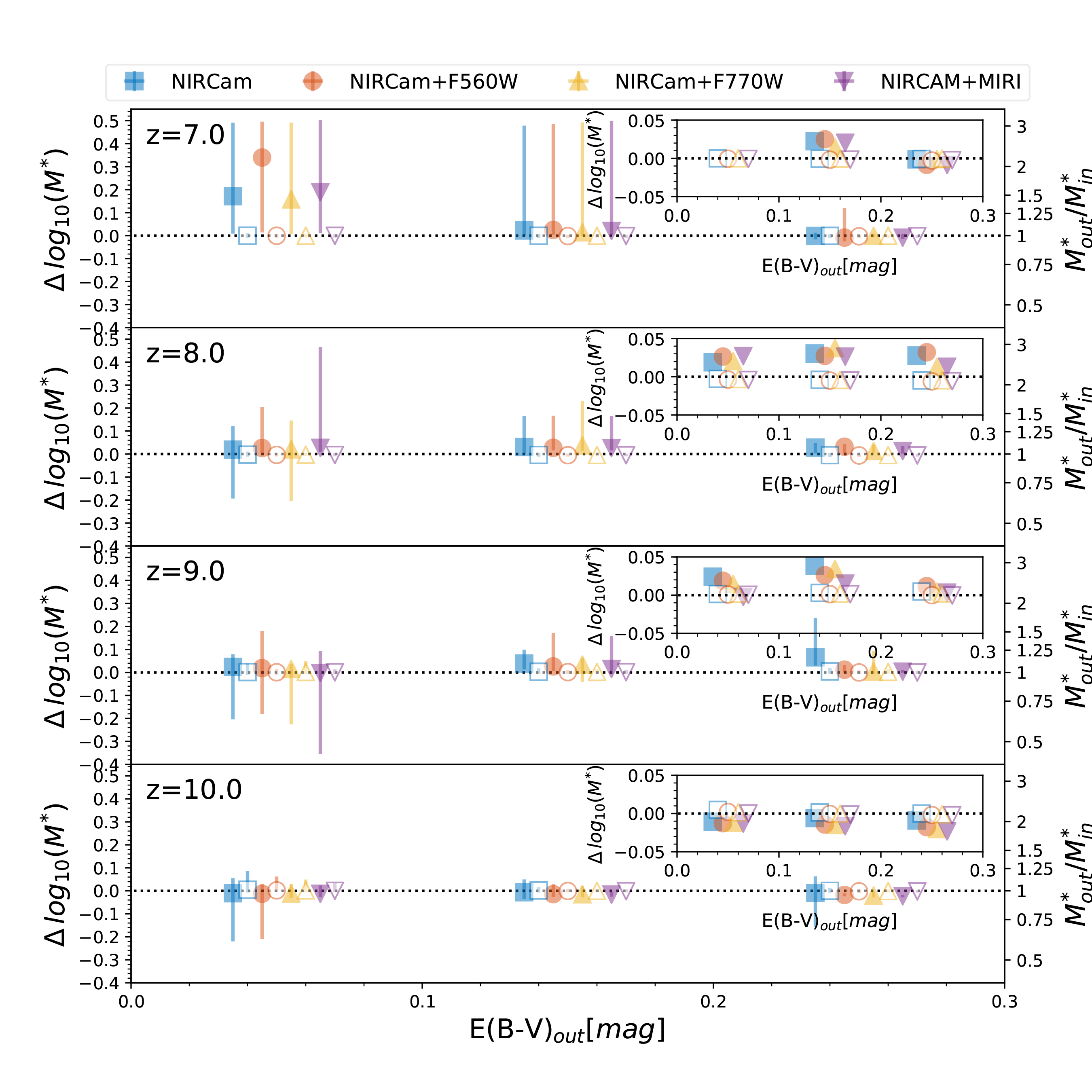}}
		\caption{Median statistical stellar mass correction for different output color excess bins and different output redshift bins. \textit{From top to bottom}: redshifts z = 7, 8, 9 and 10. Different symbols correspond to different broad-band filter combinations: 8 NIRCam broad bands (\textit{blue squares}); 8 NIRCam broad bands and MIRI F560W only (\textit{red circles}); 8 NIRCam broad bands and MIRI F770W only (\textit{yellow triangles});  8 NIRCam broad bands, MIRI F560W and MIRI F770W (\textit{purple upside-down triangles}). Filled symbols indicate galaxies with nebular emission lines, i.e. star forming and with f$_{cov}=$1, while empty symbols indicate galaxies without nebular emission lines, i.e. quiescent galaxies or star-forming galaxies with f$_{cov}=$0. Error bars correspond to the 25$\%$ and 75$\%$ quartiles. Points are slightly offset horizontally respect from each other for illustrative purposes. The inset plots show a zoom-in on the median values. \label{fig:DM-EBV}}
	\end{figure*}

	In Figure \ref{fig:DM-EBV} we show the median statistical stellar mass correction for different output color excess bins, output redshifts and filter combinations. We differentiate again between galaxies with and without nebular emission lines. Galaxies without emission lines, that could be quiescent galaxies or templates corresponding to star-forming systems with f$_{cov}=$0, have stellar mass correction quartiles within 0.05 dex, i.e. $\sim12\%$ error on the input stellar mass, for all output values of color excess and redshift. Stellar mass offsets are slightly larger, with quartiles that are anyway within 0.1 dex, for galaxies without nebular emission lines at z$=$10 and E(B-V)$\sim$0.05 mag when not including observations with the F770W MIRI band. At this high redshift, the F770W MIRI band helps to estimate the dust extinction and the stellar mass, otherwise the first one is on average underestimate and the second one is overestimated. \par
	On the other hand, for galaxies with nebular emission lines the stellar mass has a larger offset respect to the input stellar mass than for galaxies without emission lines. In particular, the dispersion of the stellar mass correction distribution increases with decreasing values of the output color excess. Moreover, the stellar mass is generally overestimated at z$=$7, particularly for galaxies with E(B-V)$<$0.1 mag for which the median statistical stellar mass correction is between 0.16 dex and 0.34 dex. Conversely, at z$>$7 the stellar mass is underestimated for a not negligible fraction of galaxies with nebular emission lines, with 25$\%$ percentiles that reach also -0.36 dex, i.e. the stellar mass is underestimated by $~56\%$. At z$>$8, the stellar mass correction has a large distribution also for galaxies with $<E(B-V)>\sim$0.25 mag, if only 8 NIRCam bands are considered. Galaxies with E(B-V)$<$0.1 mag for which the stellar mass is underestimated correspond to galaxies with input $f_{cov}=1$ ($\sim25\%$), for which the age is slightly underestimated as well as the period of star formation, or galaxies with input $f_{cov}=0$ ($\sim75\%$), for which the color excess and metallicity have been underestimated while star formation continue for a longer period than in input.
	\par
	Overall, stellar masses are generally well recovered for galaxies with output color excess around 0.2-0.3 mag, while it is less accurate for galaxies for which the best SED template is relatively dust-free. This is due to the age-extinction degeneracy for galaxies with output f$_{cov}=$0 and to a more complicated degeneracies, which involve also the duration of the star-formation and the metallicity, for galaxies with output f$_{cov}=$1.
	
	\subsection{Variation of the statistical stellar mass correction with star formation history}
\begin{figure*}[ht!]
		\center{
			\includegraphics[trim={0cm 0cm 0cm 0cm},clip,width=0.8\linewidth, keepaspectratio]{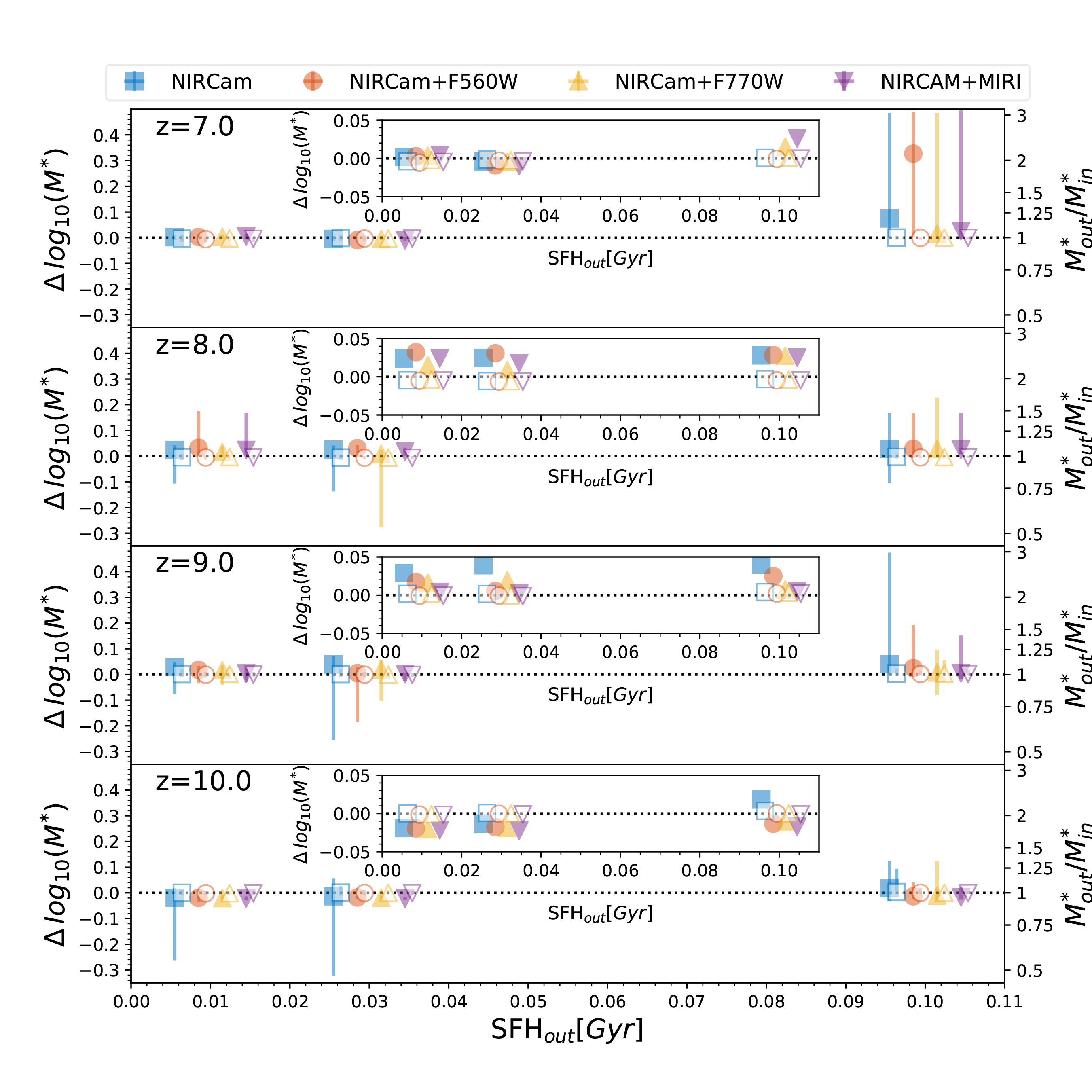}}
		\caption{Median statistical stellar mass correction for different output star formation histories and different output redshift bins. \textit{From top to bottom}: redshifts z = 7, 8, 9 and 10. Star formation histories are step functions that last for 0.01, 0.03 and 0.1 Gyr. Different symbols correspond to different broad-band filter combinations: 8 NIRCam broad bands (\textit{blue squares}); 8 NIRCam broad bands and MIRI F560W only (\textit{red circles}); 8 NIRCam broad bands and MIRI F770W only (\textit{yellow triangles});  8 NIRCam broad bands, MIRI F560W and MIRI F770W (\textit{purple upside-down triangles}). Filled symbols indicate galaxies with nebular emission lines, i.e. star forming and with f$_{cov}=$1, while empty symbols indicate galaxies without nebular emission lines, i.e. quiescent galaxies or star-forming galaxies with f$_{cov}=$0. Error bars correspond to the 25$\%$ and 75$\%$ quartiles. Points are slightly offset horizontally respect from each other for illustrative purposes.The inset plots show a zoom-in on the median values. \label{fig:DM-SFH}}
	\end{figure*}

	Figure \ref{fig:DM-SFH} shows the stellar mass corrections for different output star formation histories, redshifts and broad-band filter combinations. In this work we consider step function SFH that last for 0.01, 0.03 and 0.1 Gyr. Galaxies without emission lines have both quartiles of the stellar mass correction within 0.05 dex, excluding galaxies at z$=$10 and star formation lasting for 0.1 Gyr observed using only 8 NIRCam bands, for which the 75$\%$ quartile of the stellar mass correction is 0.09 dex. \par
	
	Galaxies with nebular emission lines have large stellar mass offset, with 75$\%$ quartiles as large as 0.5 dex, particularly for galaxies with SFH that last 0.1 Gyr at z$<$10. Indeed, some of these galaxies are correctly identified as star-forming galaxies with nebular emission lines, but the SFH and age are overestimated, with a consequent underestimation of the line equivalent widths and, therefore, overestimation of the stellar mass. On the other hand, if the galaxy was originally without emission lines but wrongly identified with f$_{cov}=$1, the age has been usually largely underestimated and, as a consequence, the stellar mass is overestimated because of a large mass-to-light ratio. If the SFH and age are underestimated the line equivalent widths is overestimated and the stellar mass, consequently, is underestimated. This is the case for galaxies with SFH shorter than 0.1 Gyr, for which the stellar mass tends to be underestimated down to -0.3 dex, i.e. around half of the input stellar mass, particularly when observed using only the 8 NIRCam bands. \par
	
	\subsection{Variation of the stellar mass correction with metallicity}
\begin{figure*}[ht!]
		\center{
			\includegraphics[trim={0cm 0cm 0cm 0cm},clip,width=0.8\linewidth, keepaspectratio]{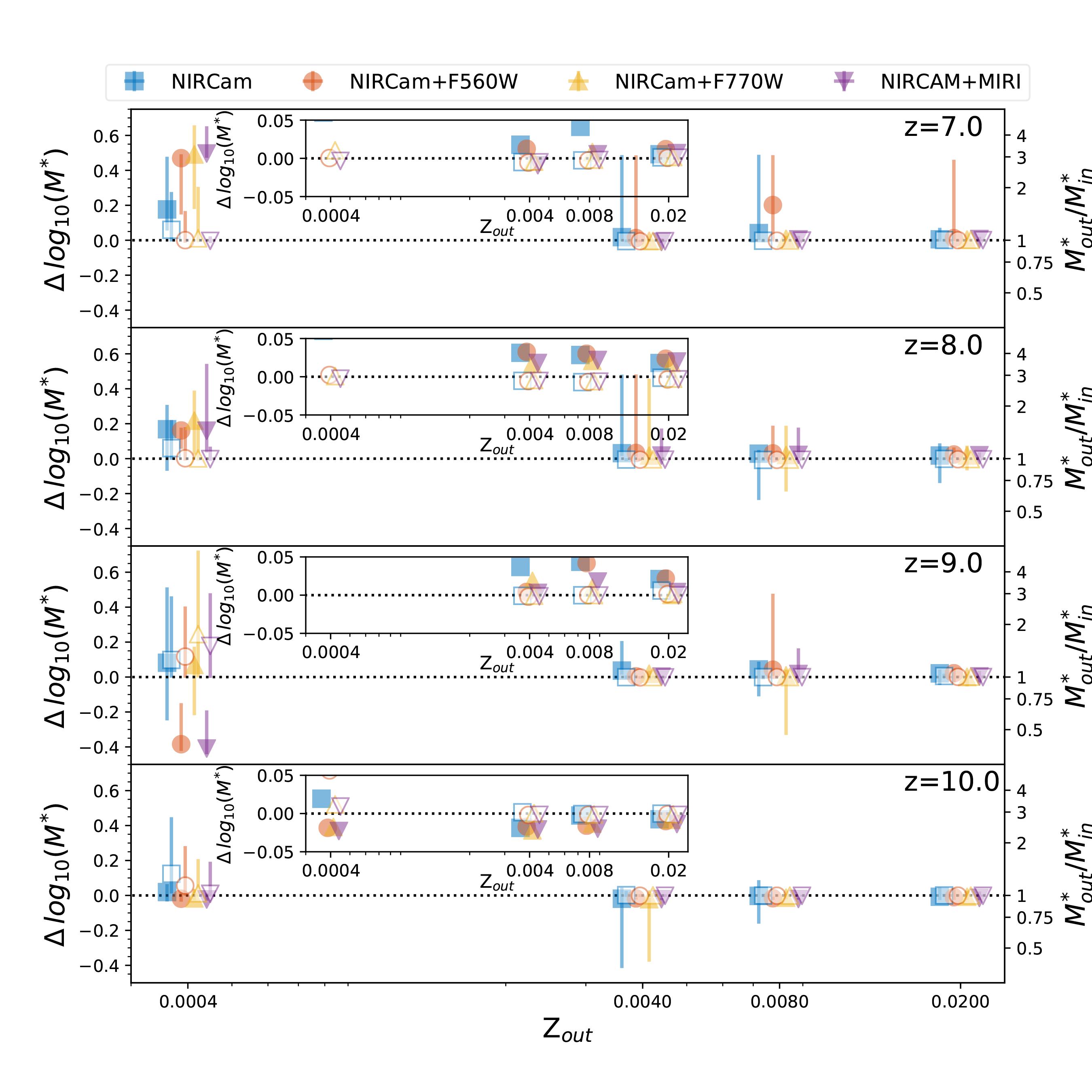}}
		\caption{Median statistical stellar mass correction for different output metallicities and different output redshift bins. \textit{From top to bottom}: redshifts z = 7, 8, 9 and 10. Different symbols correspond to different broad-band filter combinations: 8 NIRCam broad bands (\textit{blue squares}); 8 NIRCam broad bands and MIRI F560W only (\textit{red circles}); 8 NIRCam broad bands and MIRI F770W only (\textit{yellow triangles});  8 NIRCam broad bands, MIRI F560W and MIRI F770W (\textit{purple upside-down triangles}). Filled symbols indicate galaxies with nebular emission lines, i.e. star forming and with f$_{cov}=$1, while empty symbols indicate galaxies without nebular emission lines, i.e. quiescent galaxies or star-forming galaxies with f$_{cov}=$0.  Error bars correspond to the 25$\%$ and 75$\%$ quartiles. Points are slightly offset horizontally respect from each other for illustrative purposes. The inset plots show a zoom-in on the median values. \label{fig:DM-Z}}
	\end{figure*}

	In Figure \ref{fig:DM-Z}, we show the stellar mass correction for different output redshifts, output metallicities and different broad-band filter combinations. The largest scatter in the stellar mass correction is measured for galaxies with the lowest metallicity, i.e. 1$/$50 Z$_{\odot}$. Galaxies without emission lines and with metallicities larger than Z$=$0.0004 have extremely small stellar mass corrections, with both quartiles within 0.025 dex. Instead, at the lowest considered metallicity, galaxies without emission line may have stellar mass corrections with 75$\%$ quartiles as large as 0.72 dex. \par
	
	On the other hand, galaxies with nebular emission lines have large stellar mass corrections at all considered metallicities. For galaxies with nebular emission lines, stellar masses may be largely overestimated, with 75$\%$ quartiles as large as 0.66 dex, as well as largely underestimated, with 25$\%$ quartiles down to -0.44 dex. When the stellar mass is overestimated both the color excess and redshift are generally underestimated, while the age is overestimated.  For galaxies with nebular emission lines the SFH is generally underestimated and, as the age is overestimated, the equivalent-widths of the nebular emission lines are underestimated and, as a consequence, the stellar mass is overestimated. As expected, when the stellar mass is underestimated the opposite situation happens. For clarification, the redshift offset are generally $\delta$z$=|z_{in}-z_{out}|/(1+z_{in})<$0.15, as there are almost no redshift outliers in the analysed sample. \par
	
	\section{How to apply corrections to stellar masses based on \textit{JWST} observations}\label{sec:application}
	In the previous Sections, we have discussed the statistical stellar mass corrections derived from our mock galaxy catalogue, by analyzing for which templates the stellar mass estimation is particularly challenging. In this section, we give some practice information on how to include these results on the stellar mass estimation of galaxies observed with \textit{JWST} in the near future.\par
	First, we remind the reader that, under the assumption that emission lines depend on SFH, metallicity, covering factor, age and dust extinction, but not on stellar mass, the stellar mass correction can be applied to any stellar mass and does not depend on the assumed stellar mass distribution. On the other hand, we advise to use these corrections directly only for stellar masses derived using the \textit{LePhare} code and the SED template considered in this work. If other codes or templates are used or small and biased samples are considered, like samples containing only emission-line galaxies, the stellar mass corrections can be anyway used as an indication of the possible degeneracies. \par
	
	We provide the 25$\%$, 50$\%$ and 75$\%$ quartiles, minimum and maximum value of the statistical stellar mass corrections for each combination of output galaxy parameter for galaxies with f$_{cov}=$0 (Table \ref{tab:DM_noel}) and with f$_{cov}=$1 (Table \ref{tab:DM_el}) separately. We also give the entire stellar mass offset distribution for each specific SED template for both S/N$=$10 and 20 and for the four \textit{JWST} broad-band filter combinations considered: 8 NIRCam broad-bands (Table \ref{tab:PDM_1}), 8 NIRCam broad-bands and MIRI/F560W (Table \ref{tab:PDM_2}), 8 NIRCam broad-bands and MIRI/F770W (Table \ref{tab:PDM_3}), and 8 NIRCam broad-bands, MIRI/F560W and MIRI/F770W (Table \ref{tab:PDM_4}).\par
	
	In general, given a sample of galaxies observed using a specific set of \textit{JWST} filters, we suggest to derive the best SED template for each observed galaxy, using the template utilized in this paper. After obtaining the best SED template, it is possible to consult Tables \ref{tab:DM_noel} or \ref{tab:DM_el}, depending on the output covering factor, to derive the 25$\%$, 50$\%$ and 75$\%$ quartiles of the stellar mass offset, or to look at Tables \ref{tab:PDM_1} to \ref{tab:PDM_4}, depending on the available \textit{JWST} bands, to obtain the full stellar mass offset distribution. We highlight that the stellar mass offsets presented in this work need to be considered as statistical corrections, therefore their application on small samples or single objects is not recommended. They can be used on small or biased samples only to identify relatively secure or challenging galaxy templates for which the stellar mass estimation may be considered relatively secure or needs a further, more focused investigation.\par 
	
\begin{figure*}[ht!]
		\center{
			\includegraphics[trim={0cm 0cm 0cm 0cm},clip,width=0.8\linewidth, keepaspectratio]{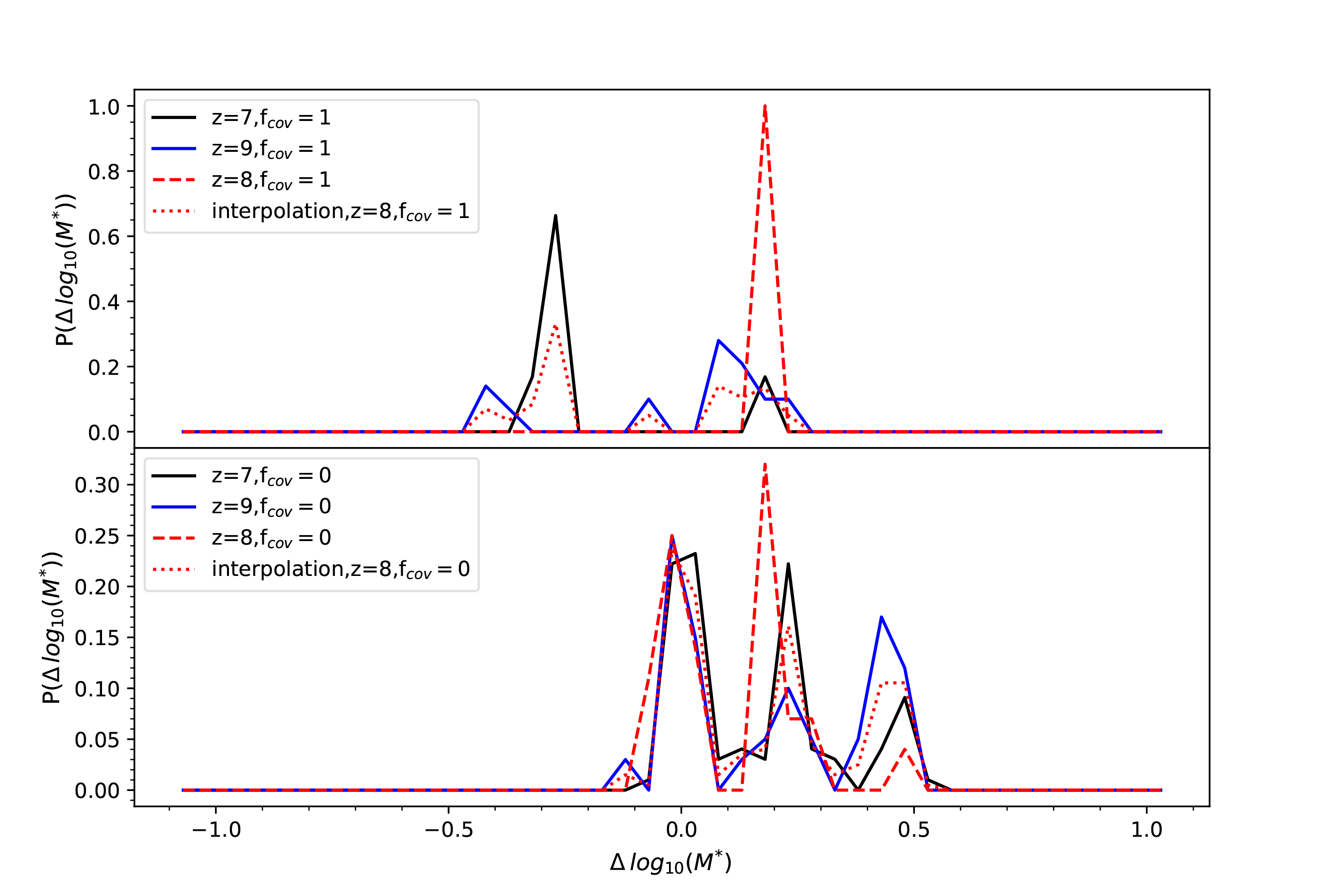}}
		\caption{Example of a distribution of the statistical stellar mass corrections for three SED templates with the same parameter values, except the redshift. We show the distribution for a template at z$=$7 (\textit{solid black line}), z$=$8 (\textit{dashed red line}), z$=$9 (\textit{solid blue line}) and for the interpolation between the distribution for the template at z$=$7 and z$=$9 (\textit{dotted red line}). We show two examples, i.e. for templates with f$_{cov}=$1 (\textit{top}) and f$_{cov}=$0 (\textit{bottom}). \label{fig:interp_z}}
	\end{figure*}

\begin{figure*}[ht!]
		\center{
			\includegraphics[trim={0cm 0cm 0cm 0cm},clip,width=0.8\linewidth, keepaspectratio]{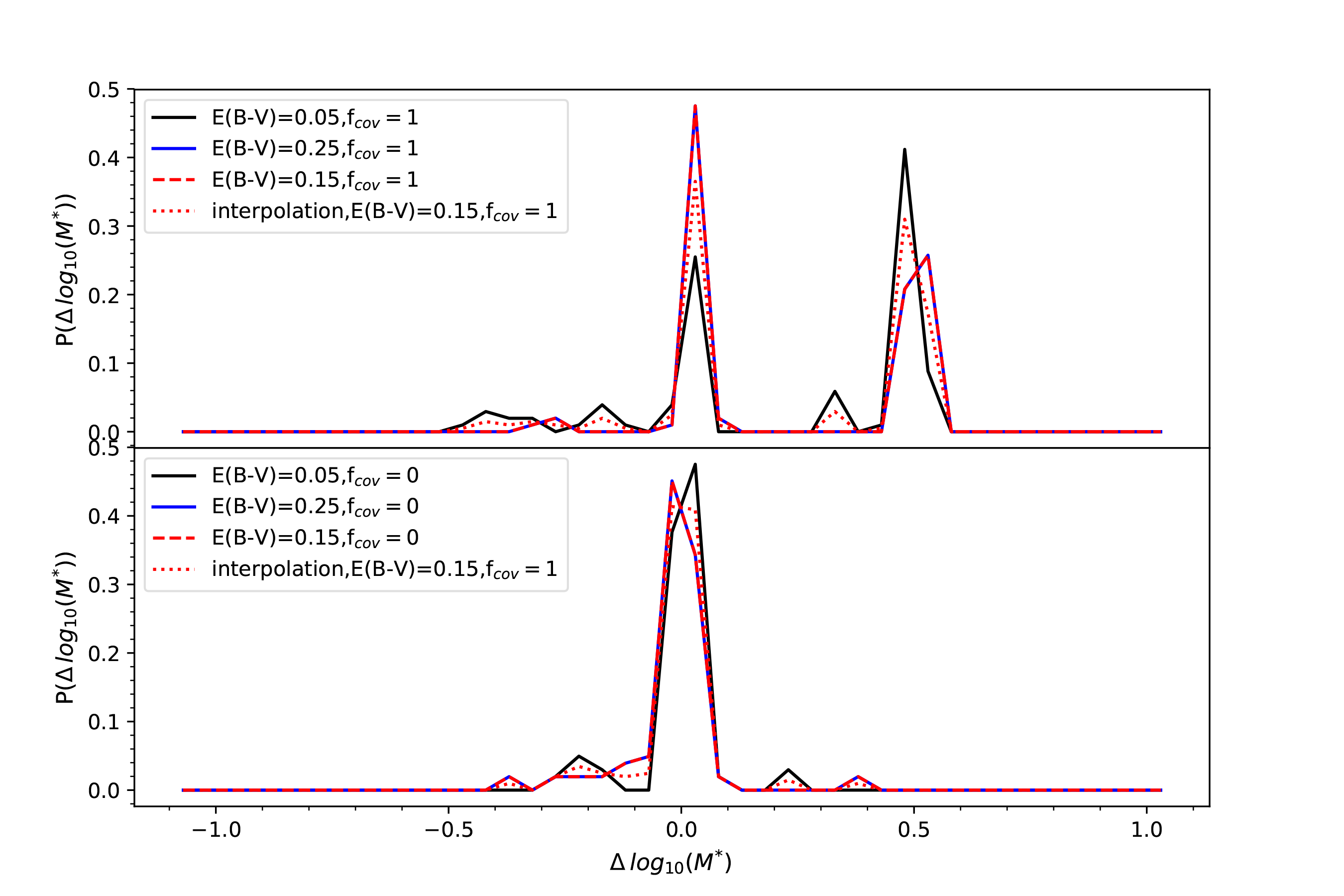}}
		\caption{Example of a distribution of the statistical stellar mass corrections for three SED templates with the same parameter values, except the color excess. We show the distribution for a template with E(B-V)$=$0.05 mag (\textit{solid black line}), E(B-V)$=$0.15 mag (\textit{dashed red line}), E(B-V)$=$0.25 mag (\textit{solid blue line}) and for the interpolation between the distribution for the template with E(B-V)$=$0.05 mag and E(B-V)$=$0.25 mag (\textit{dotted red line}). We show two examples, i.e. for templates with f$_{cov}=$1 (\textit{top}) and f$_{cov}=$0 (\textit{bottom}).\label{fig:interp_EBV}}
	\end{figure*}

	It is possible to interpolate among stellar mass offset distributions with similar output parameter values, however we suggest to perform this interpolation with caution, particularly for galaxies with emission lines. Figures \ref{fig:interp_z} and \ref{fig:interp_EBV} show some example of linear interpolation between available parameter values. In particular, in Figure \ref{fig:interp_z}, we show the interpolation between two templates with z$=$7 and 9, with all other parameters identical, and we compare the result with the corresponding template at z$=$8. In Figure \ref{fig:interp_EBV}, we repeat the exercise by changing the color excess and, in particular, by comparing the interpolation between a template with $<E(B-V)>=$0.05 mag and 0.25 mag with the corresponding template with $<$E(B-V)$>=$0.15 mag. The interpolated template never coincides perfectly with the real one, but it may resemble it, particularly for galaxies without emission lines. Therefore, we suggest to interpolate with caution among the parameter values considered in the work. \par
	
	\begin{deluxetable*}{ccccc|ccccc|ccccc|ccccc|ccccc}
		\rotate
		\tablecaption{Statistical stellar mass corrections $\Delta(log_{10}(M^*))=log_{10}(M^{*}_{out}/M_{\odot})-log_{10}(M^{*}_{in}/M_{\odot})$ for different output galaxy property combinations for galaxies with output f$_{cov}=$0. Columns from 6 to 25 show the 25$\%$, 50$\%$ and 75$\%$ quartiles, minimum and maximum values of stellar mass correction for different \textit{JWST} filter combinations. Only galaxy with Median statistical stellar mass correction larger than 0.1 dex, in at least one \textit{JWST} filter combination are shown, but the complete table with all galaxy models is available online. A value of -99.9 is present when there are no galaxies for a combination of output galaxy properties.\label{tab:DM_noel}}
		\tabletypesize{\scriptsize}
		\tablecolumns{25}
		\tablewidth{700pt}
		\tablenum{2}
		\tablehead{
			\colhead{$<$z$>$} & \colhead{Z} & \colhead{SFH} & \colhead{t} & \colhead{$<$E(B-V)$>$} & \multicolumn{5}{c}{NIRCam} & \multicolumn{5}{c}{NIRCam} & \multicolumn{5}{c}{NIRCam} & \multicolumn{5}{c}{NIRCam} \\
			\colhead{}& \colhead{[Z$_{\odot}$]} & \colhead{[Gyr]} & \colhead{[Gyr]}  & \colhead{[mag]}  & \multicolumn{5}{c}{} & \multicolumn{5}{c}{+F560W} & \multicolumn{5}{c}{+F770W} & \multicolumn{5}{c}{+F560W+F770W}\\
			\hline
			\colhead{}& \colhead{} & \colhead{} & \colhead{} & \colhead{} & \colhead{25$\%$} & \colhead{50$\%$} & \colhead{75$\%$} & \colhead{Min} & \colhead{Max} & \colhead{25$\%$} & \colhead{50$\%$} & \colhead{75$\%$} & \colhead{Min} & \colhead{Max} & \colhead{25$\%$} & \colhead{50$\%$} & \colhead{75$\%$} & \colhead{Min} & \colhead{Max} & \colhead{25$\%$} & \colhead{50$\%$} & \colhead{75$\%$} & \colhead{Min} & \colhead{Max}
		}
		\tablehead{
			\colhead{$<$z$>$} & \colhead{Z} & \colhead{SFH} & \colhead{t} & \colhead{$<$E(B-V)$>$} & \multicolumn{5}{c}{NIRCam} & \multicolumn{5}{c}{NIRCam} & \multicolumn{5}{c}{NIRCam} & \multicolumn{5}{c}{NIRCam} \\
			\colhead{}& \colhead{[Z$_{\odot}$]} & \colhead{[Gyr]} & \colhead{[Gyr]}  & \colhead{[mag]}  & \multicolumn{5}{c}{} & \multicolumn{5}{c}{+F560W} & \multicolumn{5}{c}{+F770W} & \multicolumn{5}{c}{+F560W+F770W}\\
			\hline
			\colhead{}& \colhead{} & \colhead{} & \colhead{} & \colhead{} & \colhead{25$\%$} & \colhead{50$\%$} & \colhead{75$\%$} & \colhead{Min} & \colhead{Max} & \colhead{25$\%$} & \colhead{50$\%$} & \colhead{75$\%$} & \colhead{Min} & \colhead{Max} & \colhead{25$\%$} & \colhead{50$\%$} & \colhead{75$\%$} & \colhead{Min} & \colhead{Max} & \colhead{25$\%$} & \colhead{50$\%$} & \colhead{75$\%$} & \colhead{Min} & \colhead{Max}
		}
		\startdata  
		7 & 0.02 & 0.1 & 0.05 & 0.05 & 0.32 & 0.19 & 0.72 & -0.03 & 0.78 & 0.20 & 0.00 & 0.34 & -0.03 & 0.73 & 0.44 & 0.20 & 0.72 & -0.03 & 0.78 & 0.20 & 0.00 & 0.34 & -0.03 & 0.73\\ 
		7 & 0.02 & 0.01 & 0.01 & 0.15 & 0.16 & -0.01 & 0.30 & -0.27 & 0.48 & 0.00 & -0.02 & 0.20 & -0.24 & 0.32 & 0.02 & -0.01 & 0.31 & -0.22 & 0.48 & 0.00 & -0.02 & 0.03 & -0.23 & 0.33\\ 
		7 & 0.02 & 0.03 & 0.01 & 0.15 & 0.21 & -0.01 & 0.31 & -0.07 & 0.47 & 0.18 & -0.01 & 0.21 & -0.03 & 0.31 & 0.19 & 0.00 & 0.46 & -0.03 & 0.61 & 0.00 & -0.02 & 0.19 & -0.03 & 0.31\\ 
		7& 0.02 & 0.1 & 0.01 & 0.15 & 0.19 & 0.01 & 0.28 & -0.06 & 0.49 & 0.03 & -0.01 & 0.23 & -0.05 & 0.32 & 0.18 & 0.00 & 0.31 & -0.03 & 0.61 & 0.00 & -0.02 & 0.19 & -0.03 & 0.32\\ 
		7& 0.02 & 0.1 & 0.05 & 0.15 & 0.40 & 0.20 & 0.70 & -0.08 & 0.88 & 0.01 & -0.02 & 0.19 & -0.08 & 0.89 & 0.45 & 0.21 & 0.70 & -0.08 & 0.88 & 0.00 & -0.02 & 0.08 & -0.08 & 0.89\\ 
		7& 0.02 & 0.03 & 0.01 & 0.25 & 0.12 & 0.00 & 0.23 & -0.05 & 0.51 & 0.00 & -0.01 & 0.03 & -0.05 & 0.33 & 0.00 & -0.03 & 0.02 & -0.13 & 0.59 & 0.00 & -0.02 & 0.01 & -0.12 & 0.24\\ 
		7& 0.02 & 0.1 & 0.05 & 0.25 & 0.19 & 0.00 & 0.35 & -0.04 & 0.86 & 0.00 & -0.02 & 0.04 & -0.03 & 0.42 & 0.01 & -0.01 & 0.37 & -0.04 & 0.86 & -0.01 & -0.02 & 0.01 & -0.04 & 0.87\\ 
		8& 0.02 & 0.1 & 0.05 & 0.05 & 0.27 & 0.09 & 0.45 & -0.02 & 0.74 & 0.29 & 0.01 & 0.36 & -0.02 & 0.58 & 0.31 & 0.01 & 0.46 & -0.02 & 0.74 & 0.29 & 0.00 & 0.36 & -0.02 & 0.58\\ 
		8& 0.02 & 0.01 & 0.01 & 0.15 & 0.18 & -0.01 & 0.46 & -0.27 & 0.60 & 0.18 & -0.01 & 0.20 & -0.27 & 0.33 & 0.20 & 0.00 & 0.46 & -0.28 & 0.48 & 0.01 & -0.01 & 0.20 & -0.28 & 0.30\\ 
		8& 0.02 & 0.03 & 0.01 & 0.15 & 0.20 & 0.04 & 0.39 & -0.10 & 0.59 & 0.20 & 0.18 & 0.22 & -0.27 & 0.33 & 0.23 & 0.18 & 0.46 & -0.27 & 0.59 & 0.20 & 0.19 & 0.21 & -0.03 & 0.32\\ 
		8& 0.02 & 0.1 & 0.01 & 0.15 & 0.21 & 0.00 & 0.33 & -0.07 & 0.49 & 0.20 & -0.01 & 0.22 & -0.11 & 0.33 & 0.20 & 0.13 & 0.46 & -0.02 & 0.49 & 0.18 & -0.01 & 0.20 & -0.11 & 0.34\\ 
		8& 0.02 & 0.03 & 0.05 & 0.15 & 0.10 & 0.00 & 0.12 & -0.04 & 0.35 & 0.00 & -0.01 & 0.12 & -0.11 & 0.39 & -0.01 & -0.02 & 0.00 & -0.11 & 0.46 & -0.01 & -0.02 & 0.00 & -0.10 & 0.39\\ 
		8& 0.02 & 0.1 & 0.05 & 0.15 & 0.24 & 0.00 & 0.37 & -0.08 & 0.86 & 0.00 & -0.01 & 0.17 & -0.08 & 0.54 & 0.17 & 0.00 & 0.38 & -0.08 & 0.86 & 0.00 & -0.01 & 0.16 & -0.08 & 0.88\\ 
		8& 0.02 & 0.1 & 0.05 & 0.25 & 0.17 & 0.01 & 0.33 & -0.04 & 0.75 & 0.10 & -0.01 & 0.32 & -0.05 & 0.48 & 0.00 & -0.01 & 0.02 & -0.06 & 0.39 & 0.00 & -0.01 & 0.01 & -0.06 & 0.42\\ 
		8& 1 & 0.1 & 0.05 & 0.05 & 0.24 & 0.02 & 0.34 & -0.22 & 0.54 & 0.09 & 0.01 & 0.26 & -0.18 & 0.52 & 0.05 & 0.00 & 0.23 & -0.04 & 0.54 & 0.06 & 0.00 & 0.15 & -0.04 & 0.52\\ 
		9& 0.02 & 0.01 & 0.05 & 0.05 & 0.23 & 0.01 & 0.88 & -0.07 & 0.90 & 0.10 & 0.00 & 0.41 & -0.07 & 0.91 & 0.87 & 0.00 & 0.89 & -0.07 & 0.92 & 0.01 & -0.01 & 0.87 & -0.06 & 0.91\\ 
		9& 0.02 & 0.03 & 0.05 & 0.05 & 0.10 & -0.01 & 0.79 & -0.19 & 0.98 & 0.06 & -0.01 & 0.46 & -0.12 & 0.97 & 0.48 & 0.00 & 0.94 & -0.15 & 0.97 & 0.01 & -0.01 & 0.76 & -0.20 & 0.96\\ 
		9& 0.02 & 0.1 & 0.05 & 0.05 & 0.29 & 0.02 & 0.70 & -0.06 & 0.76 & 0.30 & 0.22 & 0.72 & -0.06 & 0.76 & 0.46 & 0.24 & 0.73 & -0.06 & 0.76 & 0.47 & 0.25 & 0.73 & -0.16 & 0.76\\ 
		9& 0.02 & 0.01 & 0.01 & 0.15 & 0.01 & -0.01 & 0.42 & -0.27 & 0.49 & 0.13 & -0.01 & 0.44 & -0.30 & 0.65 & 0.20 & 0.00 & 0.46 & -0.42 & 0.65 & 0.20 & 0.00 & 0.47 & -0.30 & 0.65\\ 
		9& 0.02 & 0.03 & 0.01 & 0.15 & 0.01 & -0.01 & 0.28 & -0.13 & 0.50 & 0.11 & -0.01 & 0.25 & -0.26 & 0.50 & 0.19 & 0.00 & 0.21 & -0.27 & 0.61 & 0.20 & 0.00 & 0.27 & -0.26 & 0.63\\ 
		9& 0.02 & 0.1 & 0.01 & 0.15 & 0.01 & -0.01 & 0.24 & -0.27 & 0.49 & 0.13 & 0.00 & 0.22 & -0.27 & 0.64 & 0.22 & 0.07 & 0.48 & -0.31 & 0.65 & 0.21 & 0.01 & 0.61 & -0.26 & 0.65\\ 
		9& 0.02 & 0.1 & 0.05 & 0.15 & 0.39 & 0.17 & 0.72 & -0.12 & 0.90 & 0.41 & 0.24 & 0.73 & -0.14 & 0.91 & 0.72 & 0.26 & 0.74 & -0.26 & 0.91 & 0.71 & 0.25 & 0.87 & -0.14 & 0.92\\ 
		9& 0.02 & 0.01 & 0.01 & 0.25 & 0.00 & -0.01 & 0.28 & -0.31 & 0.49 & 0.21 & 0.00 & 0.47 & -0.27 & 0.65 & 0.00 & -0.01 & 0.20 & -0.23 & 0.62 & 0.46 & 0.00 & 0.48 & -0.22 & 0.65\\ 
		9& 0.02 & 0.03 & 0.01 & 0.25 & 0.20 & 0.00 & 0.41 & -0.11 & 0.49 & 0.16 & 0.00 & 0.40 & -0.07 & 0.66 & 0.00 & -0.01 & 0.01 & -0.07 & 0.62 & 0.00 & -0.01 & 0.02 & -0.08 & 0.66\\ 
		9& 0.02 & 0.01 & 0.05 & 0.25 & 0.39 & 0.00 & 0.87 & -0.09 & 0.90 & -0.01 & -0.01 & 0.01 & -0.07 & 0.40 & 0.86 & 0.00 & 0.87 & -0.04 & 0.89 & -0.01 & -0.01 & 0.00 & -0.03 & 0.21\\ 
		9& 0.02 & 0.1 & 0.05 & 0.25 & 0.66 & 0.25 & 0.70 & -0.13 & 0.89 & 0.23 & 0.01 & 0.38 & -0.12 & 0.89 & 0.70 & 0.26 & 0.72 & -0.08 & 0.91 & 0.39 & 0.01 & 0.70 & -0.07 & 0.90\\ 
		9& 0.2 & 0.1 & 0.05 & 0.05 & 0.19 & 0.00 & 0.34 & -0.33 & 0.73 & 0.01 & -0.01 & 0.28 & -0.32 & 0.71 & 0.19 & 0.00 & 0.49 & -0.33 & 0.73 & 0.00 & -0.01 & 0.28 & -0.32 & 0.71\\ 
		9& 0.2 & 0.03 & 0.01 & 0.15 & 0.00 & -0.08 & 0.01 & -0.46 & 0.44 & 0.00 & -0.05 & 0.02 & -0.46 & 0.45 & 0.01 & -0.01 & 0.41 & -0.53 & 0.45 & 0.40 & -0.01 & 0.42 & -0.39 & 0.45\\ 
		9& 0.4 & 0.1 & 0.05 & 0.05 & 0.22 & 0.00 & 0.48 & -0.30 & 0.70 & 0.06 & 0.00 & 0.29 & -0.32 & 0.68 & 0.19 & 0.00 & 0.48 & -0.33 & 0.69 & 0.08 & 0.00 & 0.39 & -0.32 & 0.57\\ 
		9& 1 & 0.1 & 0.05 & 0.25 & 0.34 & 0.00 & 0.41 & -0.30 & 0.48 & 0.00 & -0.03 & 0.24 & -0.31 & 0.49 & -0.01 & -0.03 & 0.01 & -0.11 & 0.54 & 0.00 & -0.02 & 0.02 & -0.13 & 0.05\\ 
		10& 0.02 & 0.03 & 0.05 & 0.05 & 0.11 & 0.01 & 0.80 & -0.13 & 0.83 & 0.09 & 0.00 & 0.35 & -0.13 & 0.98 & 0.00 & -0.01 & 0.05 & -0.13 & 0.98 & 0.00 & -0.01 & 0.07 & -0.12 & 0.98\\ 
		10& 0.02 & 0.1 & 0.05 & 0.05 & 0.34 & 0.25 & 0.56 & -0.03 & 0.77 & 0.32 & 0.22 & 0.48 & -0.16 & 0.75 & 0.32 & 0.26 & 0.47 & -0.03 & 0.76 & 0.32 & 0.22 & 0.47 & -0.03 & 0.76\\ 
		10& 0.02 & 0.01 & 0.01 & 0.15 & 0.02 & -0.02 & 0.28 & -0.28 & 0.50 & 0.07 & 0.02 & 0.28 & -0.28 & 0.48 & 0.21 & 0.04 & 0.29 & -0.29 & 0.30 & 0.21 & 0.02 & 0.29 & -0.28 & 0.30\\ 
		10& 0.02 & 0.03 & 0.01 & 0.15 & 0.29 & 0.02 & 0.30 & -0.02 & 0.48 & 0.18 & 0.01 & 0.29 & -0.27 & 0.49 & 0.28 & 0.22 & 0.29 & -0.28 & 0.30 & 0.28 & 0.22 & 0.29 & -0.27 & 0.30\\ 
		10& 0.02 & 0.1 & 0.01 & 0.15 & 0.29 & 0.07 & 0.46 & -0.02 & 0.49 & 0.19 & -0.02 & 0.21 & -0.08 & 0.46 & 0.21 & 0.20 & 0.28 & -0.08 & 0.44 & 0.20 & -0.01 & 0.23 & -0.08 & 0.44\\ 
		10& 0.02 & 0.1 & 0.05 & 0.15 & 0.51 & 0.17 & 0.71 & -0.09 & 0.89 & 0.22 & 0.01 & 0.54 & -0.07 & 0.76 & 0.20 & 0.01 & 0.48 & -0.11 & 0.57 & 0.13 & 0.01 & 0.46 & -0.10 & 0.57\\ 
		10& 0.02 & 0.03 & 0.01 & 0.25 & 0.28 & 0.20 & 0.30 & -0.08 & 0.48 & 0.20 & 0.00 & 0.22 & -0.05 & 0.30 & 0.00 & 0.00 & 0.01 & -0.03 & 0.45 & 0.00 & 0.00 & 0.01 & -0.03 & 0.43\\ 
		10& 0.02 & 0.1 & 0.01 & 0.25 & 0.29 & 0.10 & 0.30 & -0.29 & 0.45 & 0.20 & 0.01 & 0.29 & -0.06 & 0.31 & 0.00 & -0.01 & 0.02 & -0.05 & 0.42 & 0.00 & -0.01 & 0.01 & -0.04 & 0.32\\ 
		\enddata	
	\end{deluxetable*}

	\begin{deluxetable*}{ccccc|ccccc|ccccc|ccccc|ccccc}
		\rotate
		\tablecaption{ \textit{(continued)}}\label{tab:DM_noel2}
		\tabletypesize{\scriptsize}
		\tablecolumns{25}
		\tablewidth{700pt}
		\tablenum{2}
		\tablehead{
			\colhead{$<$z$>$} & \colhead{Z} & \colhead{SFH} & \colhead{t} & \colhead{$<$E(B-V)$>$} & \multicolumn{5}{c}{NIRCam} & \multicolumn{5}{c}{NIRCam} & \multicolumn{5}{c}{NIRCam} & \multicolumn{5}{c}{NIRCam} \\
			\colhead{}& \colhead{[Z$_{\odot}$]} & \colhead{[Gyr]} & \colhead{[Gyr]}  & \colhead{[mag]}  & \multicolumn{5}{c}{} & \multicolumn{5}{c}{+F560W} & \multicolumn{5}{c}{+F770W} & \multicolumn{5}{c}{+F560W+F770W}\\
			\hline
			\colhead{}& \colhead{} & \colhead{} & \colhead{} & \colhead{} & \colhead{25$\%$} & \colhead{50$\%$} & \colhead{75$\%$} & \colhead{Min} & \colhead{Max} & \colhead{25$\%$} & \colhead{50$\%$} & \colhead{75$\%$} & \colhead{Min} & \colhead{Max} & \colhead{25$\%$} & \colhead{50$\%$} & \colhead{75$\%$} & \colhead{Min} & \colhead{Max} & \colhead{25$\%$} & \colhead{50$\%$} & \colhead{75$\%$} & \colhead{Min} & \colhead{Max}
		}
		\tablehead{
			\colhead{$<$z$>$} & \colhead{Z} & \colhead{SFH} & \colhead{t} & \colhead{$<$E(B-V)$>$} & \multicolumn{5}{c}{NIRCam} & \multicolumn{5}{c}{NIRCam} & \multicolumn{5}{c}{NIRCam} & \multicolumn{5}{c}{NIRCam} \\
			\colhead{}& \colhead{[Z$_{\odot}$]} & \colhead{[Gyr]} & \colhead{[Gyr]}  & \colhead{[mag]}  & \multicolumn{5}{c}{} & \multicolumn{5}{c}{+F560W} & \multicolumn{5}{c}{+F770W} & \multicolumn{5}{c}{+F560W+F770W}\\
			\hline
			\colhead{}& \colhead{} & \colhead{} & \colhead{} & \colhead{} & \colhead{25$\%$} & \colhead{50$\%$} & \colhead{75$\%$} & \colhead{Min} & \colhead{Max} & \colhead{25$\%$} & \colhead{50$\%$} & \colhead{75$\%$} & \colhead{Min} & \colhead{Max} & \colhead{25$\%$} & \colhead{50$\%$} & \colhead{75$\%$} & \colhead{Min} & \colhead{Max} & \colhead{25$\%$} & \colhead{50$\%$} & \colhead{75$\%$} & \colhead{Min} & \colhead{Max}
		}
		\startdata  
		10& 0.02 & 0.03 & 0.05 & 0.25 & 0.12 & 0.00 & 0.82 & -0.03 & 0.84 & 0.00 & -0.01 & 0.01 & -0.04 & 0.83 & -0.01 & -0.01 & 0.00 & -0.04 & 0.38 & -0.01 & -0.01 & 0.00 & -0.04 & 0.15\\ 
		10& 0.02 & 0.1 & 0.05 & 0.25 & 0.30 & 0.20 & 0.68 & -0.06 & 0.87 & 0.27 & 0.01 & 0.33 & -0.11 & 0.76 & 0.01 & 0.00 & 0.02 & -0.11 & 0.57 & 0.00 & -0.01 & 0.01 & -0.10 & 0.56\\ 
		10& 0.2 & 0.1 & 0.05 & 0.05 & 0.28 & 0.08 & 0.52 & -0.20 & 0.73 & 0.29 & 0.03 & 0.56 & -0.16 & 0.73 & 0.28 & 0.02 & 0.58 & -0.18 & 0.73 & 0.25 & 0.01 & 0.57 & -0.20 & 0.73\\ 
		10& 0.2 & 0.1 & 0.05 & 0.15 & 0.20 & 0.00 & 0.69 & -0.25 & 0.72 & 0.02 & -0.01 & 0.23 & -0.24 & 0.72 & 0.22 & 0.00 & 0.70 & -0.24 & 0.72 & 0.01 & -0.01 & 0.24 & -0.24 & 0.72\\ 
		10& 0.2 & 0.1 & 0.05 & 0.25 & 0.28 & 0.02 & 0.68 & -0.24 & 0.72 & 0.01 & -0.01 & 0.22 & -0.23 & 0.72 & 0.69 & 0.00 & 0.71 & -0.10 & 0.73 & 0.00 & -0.01 & 0.02 & -0.09 & 0.73\\ 
		10& 0.4 & 0.1 & 0.05 & 0.05 & 0.27 & 0.08 & 0.37 & -0.27 & 0.71 & 0.25 & 0.07 & 0.37 & -0.27 & 0.71 & 0.14 & 0.04 & 0.31 & -0.27 & 0.71 & 0.13 & 0.02 & 0.29 & -0.27 & 0.71\\ 
		10& 1 & 0.1 & 0.05 & 0.05 & 0.12 & 0.05 & 0.34 & -0.23 & 0.54 & 0.12 & 0.05 & 0.33 & -0.22 & 0.55 & 0.12 & 0.05 & 0.34 & -0.23 & 0.55 & 0.11 & 0.05 & 0.27 & -0.23 & 0.55\\ 
		10& 1 & 0.03 & 0.05 & 0.15 & 0.13 & 0.01 & 0.20 & -0.09 & 0.63 & 0.07 & 0.00 & 0.19 & -0.04 & 0.66 & 0.07 & 0.00 & 0.20 & -0.10 & 0.77 & 0.00 & -0.01 & 0.06 & -0.05 & 0.66\\ 
		10& 1 & 0.1 & 0.05 & 0.25 & 0.25 & 0.00 & 0.35 & -0.25 & 0.53 & 0.01 & -0.01 & 0.27 & -0.24 & 0.54 & 0.00 & -0.01 & 0.01 & -0.06 & 0.36 & 0.00 & -0.02 & 0.01 & -0.06 & 0.33\\ 
		\enddata	
	\end{deluxetable*}
	
	\begin{deluxetable*}{ccccc|ccccc|ccccc|ccccc|ccccc} 	
		\rotate
		\tablecaption{{\footnotesize Statistical stellar mass corrections $\Delta(log_{10}(M^*))=log_{10}(M^{*}_{out}/M_{\odot})-log_{10}(M^{*}_{in}/M_{\odot})$ for different output galaxy property combinations for galaxies with output f$_{cov}=$1. Columns from 6 to 25 show the 25$\%$, 50$\%$ and 75$\%$ quartiles, minimum and maximum values of stellar mass correction for different \textit{JWST} filter combinations. Only galaxy with Median statistical stellar mass correction larger than 0.1 dex, in at least one \textit{JWST} filter combination are shown, but the complete table with all galaxy models is available online. A value of -99.9 is present when there are no galaxies for a combination of output galaxy properties.}	\label{tab:DM_el}}
		\tabletypesize{\scriptsize}
		\tablecolumns{25}
		\tablewidth{700pt}
		\tablenum{3}
		\tablehead{
			\colhead{$<$z$>$} & \colhead{Z} & \colhead{SFH} & \colhead{t} & \colhead{$<$E(B-V)$>$} & \multicolumn{5}{c}{NIRCam} & \multicolumn{5}{c}{NIRCam} & \multicolumn{5}{c}{NIRCam} & \multicolumn{5}{c}{NIRCam} \\
			\colhead{}& \colhead{[Z$_{\odot}$]} & \colhead{[Gyr]} & \colhead{[Gyr]}  & \colhead{[mag]}  & \multicolumn{5}{c}{} & \multicolumn{5}{c}{+F560W} & \multicolumn{5}{c}{+F770W} & \multicolumn{5}{c}{+F560W+F770W}\\
			\hline
			\colhead{}& \colhead{} & \colhead{} & \colhead{} & \colhead{} & \colhead{25$\%$} & \colhead{50$\%$} & \colhead{75$\%$} & \colhead{Min} & \colhead{Max} & \colhead{25$\%$} & \colhead{50$\%$} & \colhead{75$\%$} & \colhead{Min} & \colhead{Max} & \colhead{25$\%$} & \colhead{50$\%$} & \colhead{75$\%$} & \colhead{Min} & \colhead{Max} & \colhead{25$\%$} & \colhead{50$\%$} & \colhead{75$\%$} & \colhead{Min} & \colhead{Max}
		}
		\startdata  
		7 & 0.02 & 0.03 & 0.01 & 0.05 & -99.9 & -99.9 & -99.9 & -99.9 & -99.9 & 0.00 & 0.00 & 0.00 & 0.00 & 0.00 & 0.17 & 0.17 & 0.18 & 0.17 & 0.19 & 0.19 & 0.19 & 0.19 & 0.19 & 0.19 \\ 
		7 & 0.02 & 0.1 & 0.01 & 0.05 & 0.17 & 0.16 & 0.17 & 0.15 & 0.18 & 0.16 & 0.16 & 0.16 & 0.15 & 0.17 & 0.17 & 0.16 & 0.17 & 0.15 & 0.20 & 0.16 & 0.16 & 0.17 & 0.15 & 0.18 \\ 
		7 & 0.02 & 0.1 & 0.05 & 0.05 & 0.48 & 0.18 & 0.49 & -0.05 & 0.67 & 0.48 & 0.18 & 0.49 & -0.06 & 0.67 & 0.66 & 0.49 & 0.66 & -0.05 & 0.68 & 0.65 & 0.49 & 0.66 & -0.25 & 0.68 \\ 
		7 & 0.02 & 0.03 & 0.01 & 0.15 & -99.9 & -99.9 & -99.9 & -99.9 & -99.9 & 0.00 & 0.00 & 0.00 & 0.00 & 0.00 & 0.34 & 0.34 & 0.34 & 0.34 & 0.34 & 0.00 & 0.00 & 0.00 & 0.00 & 0.00 \\ 
		7 & 0.02 & 0.1 & 0.01 & 0.15 & -99.9 & -99.9 & -99.9 & -99.9 & -99.9 & 0.00 & 0.00 & 0.00 & 0.00 & 0.00 & 0.34 & 0.33 & 0.35 & 0.07 & 0.36 & 0.07 & 0.07 & 0.07 & 0.07 & 0.07 \\ 
		7 & 0.02 & 0.1 & 0.05 & 0.15 & 0.48 & 0.06 & 0.50 & -0.01 & 0.67 & 0.33 & 0.03 & 0.50 & -0.01 & 0.65 & 0.65 & 0.50 & 0.67 & 0.01 & 0.69 & 0.50 & 0.49 & 0.65 & -0.15 & 0.68 \\ 
		7 & 0.02 & 0.01 & 0.01 & 0.25 & -99.9 & -99.9 & -99.9 & -99.9 & -99.9 & 0.00 & 0.00 & 0.00 & 0.00 & 0.00 & 0.18 & 0.18 & 0.18 & 0.17 & 0.18 & 0.04 & 0.04 & 0.04 & 0.04 & 0.04 \\ 
		7 & 0.02 & 0.03 & 0.01 & 0.25 & -0.30 & -0.3 & -0.30 & -0.33 & 0.18 & 0.00 & 0.00 & 0.00 & 0.00 & 0.00 & 0.18 & 0.18 & 0.19 & 0.04 & 0.20 & 0.04 & 0.04 & 0.04 & 0.02 & 0.04 \\ 
		7 & 0.02 & 0.1 & 0.01 & 0.25 & 0.19 & -0.32 & 0.19 & -0.33 & 0.37 & 0.04 & 0.04 & 0.04 & 0.04 & 0.04 & 0.19 & 0.18 & 0.35 & -0.12 & 0.38 & 0.03 & 0.02 & 0.03 & 0.01 & 0.04 \\ 
		7 & 0.02 & 0.1 & 0.05 & 0.25 & 0.00 & 0.00 & 0.47 & -0.27 & 0.49 & 0.47 & 0.00 & 0.48 & -0.02 & 0.50 & 0.01 & 0.00 & 0.48 & -0.27 & 0.50 & 0.47 & 0.00 & 0.48 & -0.02 & 0.50 \\ 
		7 & 0.2 & 0.03 & 0.01 & 0.05 & -99.9 & -99.9 & -99.9 & -99.9 & -99.9 & 0.00 & 0.00 & 0.00 & 0.00 & 0.00 & 0.00 & 0.00 & 0.00 & 0.00 & 0.00 & 0.02 & 0.02 & 0.02 & 0.02 & 0.02 \\ 
		7 & 0.2 & 0.1 & 0.01 & 0.05 & 0.18 & 0.18 & 0.18 & 0.01 & 0.19 & 0.01 & 0.01 & 0.01 & 0.01 & 0.01 & 0.01 & 0.01 & 0.01 & 0.01 & 0.02 & 0.01 & 0.00 & 0.02 & 0.00 & 0.02 \\ 
		7 & 0.2 & 0.1 & 0.05 & 0.05 & 0.49 & 0.49 & 0.50 & -0.24 & 0.66 & 0.49 & 0.49 & 0.50 & -0.16 & 0.51 & 0.49 & 0.11 & 0.49 & 0.01 & 0.50 & 0.49 & 0.47 & 0.49 & -0.01 & 0.50 \\ 
		7 & 0.2 & 0.1 & 0.01 & 0.15 & -99.9 & -99.9 & -99.9 & -99.9 & -99.9 & 0.00 & 0.00 & 0.00 & 0.00 & 0.00 & 0.00 & 0.00 & 0.00 & 0.00 & 0.00 & 0.03 & 0.03 & 0.03 & 0.03 & 0.03 \\ 
		7 & 0.2 & 0.1 & 0.05 & 0.15 & 0.48 & 0.02 & 0.49 & -0.47 & 0.50 & 0.48 & 0.01 & 0.48 & -0.39 & 0.50 & 0.00 & -0.0 & 0.01 & -0.02 & 0.49 & 0.00 & -0.01 & 0.47 & -0.04 & 0.49 \\ 
		7 & 0.2 & 0.01 & 0.01 & 0.25 & -99.9 & -99.9 & -99.9 & -99.9 & -99.9 & -0.86 & -0.86 & -0.86 & -0.87 & -0.85 & 0.00 & 0.00 & 0.00 & 0.00 & 0.00 & -0.02 & -0.02 & -0.02 & -0.02 & -0.01 \\ 
		7 & 0.2 & 0.1 & 0.01 & 0.25 & -0.01 & -0.48 & 0.00 & -0.66 & 0.02 & -0.48 & -0.49 & -0.01 & -0.67 & 0.01 & -0.01 & -0.01 & 0.00 & -0.66 & 0.02 & -0.01 & -0.17 & -0.01 & -0.66 & 0.01 \\ 
		7 & 0.2 & 0.1 & 0.05 & 0.25 & 0.49 & 0.49 & 0.49 & -0.01 & 0.50 & 0.00 & -0.01 & 0.23 & -0.01 & 0.32 & 0.00 & -0.01 & 0.00 & -0.02 & 0.33 & -0.01 & -0.01 & 0.00 & -0.02 & 0.33 \\ 
		7 & 0.4 & 0.03 & 0.01 & 0.05 & 0.19 & 0.19 & 0.19 & 0.19 & 0.19 & 0.00 & 0.00 & 0.00 & 0.00 & 0.00 & 0.00 & 0.00 & 0.00 & 0.00 & 0.00 & -0.06 & -0.18 & 0.07 & -0.30 & 0.19 \\ 
		7 & 0.4 & 0.1 & 0.01 & 0.05 & 0.19 & 0.18 & 0.20 & -0.44 & 0.21 & 0.02 & 0.02 & 0.03 & 0.01 & 0.19 & 0.17 & 0.02 & 0.19 & -0.44 & 0.20 & 0.02 & 0.01 & 0.03 & 0.01 & 0.19 \\ 
		7 & 0.4 & 0.1 & 0.05 & 0.05 & 0.49 & 0.33 & 0.50 & -0.27 & 0.69 & 0.48 & 0.33 & 0.50 & -0.25 & 0.68 & 0.48 & 0.02 & 0.50 & -0.22 & 0.52 & 0.33 & 0.02 & 0.49 & -0.25 & 0.52 \\ 
		7 & 0.4 & 0.01 & 0.01 & 0.15 & -99.9 & -99.9 & -99.9 & -99.9 & -99.9 & 0.01 & 0.00 & 0.02 & -0.01 & 0.02 & 0.00 & 0.00 & 0.00 & 0.00 & 0.00 & 0.01 & -0.01 & 0.01 & -0.01 & 0.02 \\ 
		7 & 0.4 & 0.03 & 0.01 & 0.15 & -0.36 & -0.43 & -0.08 & -0.46 & 0.17 & 0.01 & -0.01 & 0.02 & -0.01 & 0.02 & -0.01 & -0.30 & 0.01 & -0.44 & 0.02 & 0.01 & 0.01 & 0.02 & -0.01 & 0.02 \\ 
		7 & 0.4 & 0.1 & 0.05 & 0.15 & 0.48 & 0.00 & 0.50 & -0.42 & 0.70 & 0.48 & 0.01 & 0.50 & -0.02 & 0.53 & 0.00 & 0.00 & 0.02 & -0.28 & 0.51 & 0.01 & 0.00 & 0.47 & -0.03 & 0.52 \\ 
		7 & 0.4 & 0.01 & 0.01 & 0.25 & -0.74 & -0.78 & -0.66 & -0.82 & -0.47 & -0.01 & -0.47 & -0.01 & -0.48 & 0.00 & -0.01 & -0.01 & -0.01 & -0.01 & -0.01 & -0.01 & -0.01 & -0.01 & -0.63 & 0.01 \\ 
		7 & 0.4 & 0.1 & 0.05 & 0.25 & 0.31 & -0.04 & 0.33 & -0.16 & 0.53 & 0.32 & -0.02 & 0.33 & -0.16 & 0.48 & 0.00 & -0.05 & 0.01 & -0.08 & 0.04 & 0.00 & -0.01 & 0.00 & -0.03 & 0.35 \\ 
		8 & 0.02 & 0.01 & 0.01 & 0.05 & -0.45 & -0.45 & -0.40 & -0.48 & 0.17 & 0.16 & -0.41 & 0.16 & -0.45 & 0.17 & 0.24 & 0.19 & 0.25 & -0.40 & 0.26 & 0.15 & 0.15 & 0.16 & -0.40 & 0.24 \\ 
		8 & 0.02 & 0.03 & 0.01 & 0.05 & -0.45 & -0.47 & -0.40 & -0.47 & -0.18 & -0.17 & -0.40 & -0.11 & -0.46 & 0.18 & -0.38 & -0.41 & 0.21 & -0.47 & 0.24 & -0.10 & -0.38 & 0.11 & -0.40 & 0.16 \\ 
		8 & 0.02 & 0.1 & 0.01 & 0.05 & -0.01 & -0.26 & 0.20 & -0.54 & 0.25 & -0.17 & -0.40 & 0.08 & -0.47 & 0.26 & 0.20 & -0.19 & 0.21 & -0.64 & 0.26 & 0.08 & -0.38 & 0.12 & -0.63 & 0.29 \\ 
		8 & 0.02 & 0.1 & 0.05 & 0.05 & 0.13 & -0.03 & 0.21 & -0.20 & 0.71 & 0.20 & 0.07 & 0.54 & -0.08 & 0.66 & 0.21 & 0.00 & 0.55 & -0.20 & 0.71 & 0.54 & 0.50 & 0.64 & -0.06 & 0.66 \\ 
		8 & 0.02 & 0.01 & 0.01 & 0.15 & -99.9 & -99.9 & -99.9 & -99.9 & -99.9 & 0.16 & 0.15 & 0.17 & 0.15 & 0.18 & 0.00 & 0.00 & 0.00 & 0.00 & 0.00 & 0.16 & 0.05 & 0.17 & 0.01 & 0.17 \\ 
		8 & 0.02 & 0.03 & 0.01 & 0.15 & 0.00 & 0.00 & 0.00 & 0.00 & 0.00 & 0.16 & 0.16 & 0.17 & 0.15 & 0.17 & 0.36 & 0.36 & 0.38 & 0.28 & 0.38 & 0.17 & 0.16 & 0.17 & 0.10 & 0.17 \\ 
		8 & 0.02 & 0.1 & 0.01 & 0.15 & 0.23 & -0.10 & 0.37 & -0.61 & 0.39 & 0.16 & -0.40 & 0.16 & -0.61 & 0.18 & 0.27 & -0.01 & 0.37 & -0.40 & 0.39 & 0.16 & 0.10 & 0.17 & -0.60 & 0.33 \\ 
		8 & 0.02 & 0.1 & 0.05 & 0.15 & 0.17 & 0.00 & 0.50 & -0.14 & 0.51 & 0.16 & 0.13 & 0.44 & -0.11 & 0.66 & 0.50 & 0.33 & 0.57 & -0.14 & 0.67 & 0.61 & 0.49 & 0.65 & -0.22 & 0.66 \\ 
		8 & 0.02 & 0.01 & 0.01 & 0.25 & -99.9 & -99.9 & -99.9 & -99.9 & -99.9 & 0.17 & 0.17 & 0.17 & 0.16 & 0.18 & 0.00 & 0.00 & 0.00 & 0.00 & 0.00 & 0.04 & 0.03 & 0.04 & 0.03 & 0.05 \\ 
		8 & 0.02 & 0.03 & 0.01 & 0.25 & -99.9 & -99.9 & -99.9 & -99.9 & -99.9 & 0.17 & 0.17 & 0.17 & 0.03 & 0.18 & 0.00 & 0.00 & 0.00 & 0.00 & 0.00 & 0.04 & 0.03 & 0.04 & 0.03 & 0.17 \\ 
		8 & 0.02 & 0.1 & 0.01 & 0.25 & 0.18 & 0.18 & 0.19 & 0.15 & 0.38 & 0.17 & 0.16 & 0.17 & -0.12 & 0.18 & 0.39 & 0.37 & 0.39 & 0.36 & 0.40 & 0.04 & 0.03 & 0.17 & -0.29 & 0.38 \\ 
		8 & 0.02 & 0.1 & 0.05 & 0.25 & 0.50 & 0.05 & 0.51 & -0.22 & 0.53 & 0.13 & 0.03 & 0.17 & -0.22 & 0.67 & 0.06 & 0.02 & 0.64 & -0.22 & 0.67 & 0.04 & 0.03 & 0.67 & -0.22 & 0.68 \\ 
		8 & 0.2 & 0.1 & 0.01 & 0.05 & 0.00 & -0.16 & 0.01 & -0.64 & 0.04 & -0.78 & -0.79 & -0.64 & -0.88 & -0.57 & 0.01 & -0.30 & 0.02 & -0.80 & 0.22 & 0.00 & -0.78 & 0.01 & -0.88 & 0.09 \\ 
		8 & 0.2 & 0.1 & 0.05 & 0.05 & 0.46 & 0.01 & 0.49 & -0.45 & 0.51 & 0.48 & 0.02 & 0.49 & -0.46 & 0.51 & 0.34 & 0.01 & 0.49 & -0.45 & 0.51 & 0.48 & 0.02 & 0.49 & -0.46 & 0.51 \\ 
		8 & 0.2 & 0.1 & 0.01 & 0.15 & 0.00 & -0.68 & 0.02 & -0.76 & 0.03 & -0.68 & -0.69 & -0.67 & -0.76 & -0.48 & 0.03 & 0.02 & 0.03 & -0.46 & 0.04 & 0.02 & 0.02 & 0.02 & -0.30 & 0.03 \\ 
		8 & 0.2 & 0.1 & 0.05 & 0.15 & 0.48 & 0.02 & 0.49 & -0.32 & 0.52 & 0.48 & 0.02 & 0.50 & -0.37 & 0.51 & 0.48 & 0.03 & 0.49 & -0.30 & 0.51 & 0.48 & 0.02 & 0.50 & -0.31 & 0.51 \\ 
		8 & 0.4 & 0.03 & 0.01 & 0.05 & 0.01 & -0.30 & 0.03 & -0.74 & 0.23 & 0.01 & -0.28 & 0.03 & -0.76 & 0.21 & -0.27 & -0.34 & 0.03 & -0.74 & 0.20 & 0.02 & -0.28 & 0.03 & -0.76 & 0.21 \\ 
		8 & 0.4 & 0.1 & 0.01 & 0.05 & -0.32 & -0.40 & 0.03 & -0.70 & 0.24 & 0.01 & -0.31 & 0.03 & -0.75 & 0.21 & -0.31 & -0.36 & 0.02 & -0.77 & 0.19 & 0.01 & -0.30 & 0.03 & -0.76 & 0.21 \\ 
		8 & 0.4 & 0.1 & 0.05 & 0.05 & 0.03 & -0.15 & 0.45 & -0.30 & 0.53 & 0.49 & 0.03 & 0.50 & -0.36 & 0.53 & 0.1 & 0.00 & 0.49 & -0.38 & 0.53 & 0.48 & 0.02 & 0.49 & -0.38 & 0.53 \\ 
		\enddata	
	\end{deluxetable*}
	
	\begin{deluxetable*}{ccccc|ccccc|ccccc|ccccc|ccccc} 	
		\rotate
		\tablecaption{ \textit{(continued)}}\label{tab:DM_el2}
		\tabletypesize{\scriptsize}
		\tablecolumns{25}
		\tablewidth{700pt}
		\tablenum{3}
		\tablehead{
			\colhead{$<$z$>$} & \colhead{Z} & \colhead{SFH} & \colhead{t} & \colhead{$<$E(B-V)$>$} & \multicolumn{5}{c}{NIRCam} & \multicolumn{5}{c}{NIRCam} & \multicolumn{5}{c}{NIRCam} & \multicolumn{5}{c}{NIRCam} \\
			\colhead{}& \colhead{[Z$_{\odot}$]} & \colhead{[Gyr]} & \colhead{[Gyr]}  & \colhead{[mag]}  & \multicolumn{5}{c}{} & \multicolumn{5}{c}{+F560W} & \multicolumn{5}{c}{+F770W} & \multicolumn{5}{c}{+F560W+F770W}\\
			\hline
			\colhead{}& \colhead{} & \colhead{} & \colhead{} & \colhead{} & \colhead{25$\%$} & \colhead{50$\%$} & \colhead{75$\%$} & \colhead{Min} & \colhead{Max} & \colhead{25$\%$} & \colhead{50$\%$} & \colhead{75$\%$} & \colhead{Min} & \colhead{Max} & \colhead{25$\%$} & \colhead{50$\%$} & \colhead{75$\%$} & \colhead{Min} & \colhead{Max} & \colhead{25$\%$} & \colhead{50$\%$} & \colhead{75$\%$} & \colhead{Min} & \colhead{Max}
		}
		\startdata  
		8 & 0.4 & 0.01 & 0.01 & 0.15 & 0.04 & 0.03 & 0.18 & -0.67 & 0.20 & 0.17 & 0.03 & 0.20 & -0.65 & 0.22 & 0.05 & 0.03 & 0.19 & -0.44 & 0.20 & 0.04 & 0.02 & 0.20 & -0.45 & 0.22 \\ 
		8 & 0.4 & 0.1 & 0.01 & 0.15 & -0.28 & -0.43 & 0.03 & -0.78 & 0.20 & 0.02 & -0.30 & 0.03 & -0.75 & 0.21 & -0.33 & -0.43 & -0.26 & -0.77 & 0.19 & 0.02 & -0.43 & 0.03 & -0.76 & 0.21 \\ 
		8 & 0.4 & 0.1 & 0.05 & 0.15 & 0.02 & 0.01 & 0.48 & -0.37 & 0.51 & 0.49 & 0.02 & 0.50 & -0.35 & 0.53 & 0.48 & 0.02 & 0.49 & -0.37 & 0.51 & 0.48 & 0.02 & 0.50 & -0.35 & 0.52 \\ 
		8 & 0.4 & 0.01 & 0.01 & 0.25 & 0.05 & 0.04 & 0.16 & -0.30 & 0.22 & 0.21 & 0.04 & 0.22 & -0.63 & 0.22 & 0.01 & 0.01 & 0.19 & -0.31 & 0.21 & 0.02 & 0.00 & 0.20 & -0.01 & 0.22 \\ 
		8 & 0.4 & 0.1 & 0.05 & 0.25 & 0.02 & 0.00 & 0.05 & -0.28 & 0.53 & 0.00 & 0.00 & 0.03 & -0.22 & 0.50 & 0.49 & 0.01 & 0.50 & -0.20 & 0.52 & 0.01 & 0.00 & 0.02 & -0.01 & 0.51 \\ 
		9 & 0.02 & 0.01 & 0.01 & 0.05 & -0.22 & -0.48 & 0.04 & -0.67 & 0.22 & -0.42 & -0.46 & -0.13 & -0.51 & -0.10 & -0.15 & -0.46 & 0.04 & -0.67 & 0.20 & -0.43 & -0.47 & -0.15 & -0.48 & -0.12 \\ 
		9 & 0.02 & 0.03 & 0.01 & 0.05 & -0.16 & -0.43 & 0.04 & -0.51 & 0.20 & -0.42 & -0.46 & -0.23 & -0.51 & -0.11 & -0.21 & -0.43 & 0.03 & -0.50 & 0.20 & -0.45 & -0.47 & -0.40 & -0.51 & -0.13 \\ 
		9 & 0.02 & 0.1 & 0.01 & 0.05 & -0.34 & -0.43 & 0.03 & -0.67 & 0.11 & -0.41 & -0.43 & -0.21 & -0.66 & -0.07 & -0.41 & -0.44 & 0.03 & -0.69 & 0.19 & -0.41 & -0.44 & -0.22 & -0.66 & -0.07 \\ 
		9 & 0.02 & 0.1 & 0.05 & 0.05 & 0.50 & 0.10 & 0.51 & -0.22 & 0.54 & 0.03 & -0.04 & 0.17 & -0.30 & 0.51 & 0.50 & 0.16 & 0.51 & -0.22 & 0.54 & 0.20 & 0.04 & 0.49 & -0.22 & 0.51 \\ 
		9 & 0.02 & 0.01 & 0.01 & 0.15 & 0.05 & -0.38 & 0.06 & -0.49 & 0.21 & -0.43 & -0.43 & -0.39 & -0.45 & -0.38 & 0.06 & -0.31 & 0.06 & -0.47 & 0.18 & -0.44 & -0.44 & -0.43 & -0.46 & -0.42 \\ 
		9 & 0.02 & 0.03 & 0.01 & 0.15 & -0.41 & -0.48 & 0.08 & -0.50 & 0.20 & -0.39 & -0.43 & -0.38 & -0.44 & -0.38 & 0.05 & -0.44 & 0.06 & -0.48 & 0.08 & -0.44 & -0.45 & -0.44 & -0.47 & -0.44 \\ 
		9 & 0.02 & 0.1 & 0.01 & 0.15 & -0.37 & -0.41 & 0.06 & -0.49 & 0.21 & -0.38 & -0.39 & -0.38 & -0.44 & -0.36 & 0.05 & -0.42 & 0.06 & -0.48 & 0.21 & -0.44 & -0.45 & -0.43 & -0.47 & -0.41 \\ 
		9 & 0.02 & 0.1 & 0.05 & 0.15 & 0.52 & 0.25 & 0.53 & -0.03 & 0.55 & 0.20 & -0.03 & 0.43 & -0.09 & 0.44 & 0.52 & 0.51 & 0.53 & -0.05 & 0.68 & 0.05 & -0.03 & 0.16 & -0.09 & 0.43 \\ 
		9 & 0.02 & 0.01 & 0.01 & 0.25 & 0.09 & -0.07 & 0.10 & -0.42 & 0.11 & -0.40 & -0.4 & -0.40 & -0.41 & -0.17 & 0.09 & 0.08 & 0.10 & -0.42 & 0.25 & -0.40 & -0.41 & -0.40 & -0.41 & -0.39 \\ 
		9 & 0.02 & 0.03 & 0.01 & 0.25 & 0.09 & -0.07 & 0.10 & -0.41 & 0.22 & -0.40 & -0.4 & -0.17 & -0.41 & -0.09 & 0.09 & 0.09 & 0.10 & -0.64 & 0.26 & -0.40 & -0.4 & -0.39 & -0.41 & -0.11 \\ 
		9 & 0.02 & 0.1 & 0.01 & 0.25 & -99.9 & -99.9 & -99.9 & -99.9 & -99.9 & -0.16 & -0.17 & -0.09 & -0.62 & -0.07 & 0.09 & 0.07 & 0.10 & -0.59 & 0.25 & 0.00 & 0.00 & 0.00 & 0.00 & 0.00 \\ 
		9 & 0.02 & 0.1 & 0.05 & 0.25 & 0.54 & 0.53 & 0.55 & -0.21 & 0.58 & 0.00 & -0.07 & 0.01 & -0.20 & 0.01 & 0.53 & 0.07 & 0.54 & 0.03 & 0.71 & 0.00 & 0.00 & 0.01 & -0.21 & 0.01 \\ 
		9 & 0.2 & 0.01 & 0.01 & 0.05 & -99.9 & -99.9 & -99.9 & -99.9 & -99.9 & 0.00 & 0.00 & 0.00 & 0.00 & 0.00 & 0.03 & 0.01 & 0.03 & -0.69 & 0.06 & 0.00 & 0.00 & 0.00 & 0.00 & 0.00 \\ 
		9 & 0.2 & 0.03 & 0.01 & 0.05 & -99.9 & -99.9 & -99.9 & -99.9 & -99.9 & 0.00 & 0.00 & 0.00 & 0.00 & 0.00 & 0.01 & 0.00 & 0.03 & -0.68 & 0.20 & 0.00 & 0.00 & 0.00 & 0.00 & 0.00 \\ 
		9 & 0.2 & 0.1 & 0.01 & 0.05 & 0.01 & -0.44 & 0.03 & -0.85 & 0.20 & -0.83 & -0.83 & -0.66 & -0.84 & -0.48 & 0.01 & -0.39 & 0.03 & -0.87 & 0.06 & -0.48 & -0.83 & -0.44 & -0.84 & -0.29 \\ 
		9 & 0.2 & 0.1 & 0.05 & 0.05 & 0.48 & 0.02 & 0.50 & -0.47 & 0.53 & -0.21 & -0.34 & 0.47 & -0.47 & 0.49 & 0.03 & 0.01 & 0.36 & -0.45 & 0.51 & 0.45 & -0.17 & 0.47 & -0.40 & 0.49 \\ 
		9 & 0.2 & 0.01 & 0.01 & 0.15 & 0.04 & 0.03 & 0.05 & -0.47 & 0.22 & -0.17 & -0.33 & -0.01 & -0.33 & 0.00 & 0.04 & 0.04 & 0.05 & -0.47 & 0.07 & 0.00 & -0.09 & 0.00 & -0.32 & 0.00 \\ 
		9 & 0.2 & 0.1 & 0.01 & 0.15 & 0.02 & -0.42 & 0.03 & -0.81 & 0.20 & -0.27 & -0.55 & 0.00 & -0.55 & 0.01 & 0.02 & -0.45 & 0.04 & -0.80 & 0.07 & 0.00 & -0.0 & 0.01 & -0.01 & 0.01 \\ 
		9 & 0.2 & 0.1 & 0.05 & 0.15 & 0.49 & 0.03 & 0.51 & -0.25 & 0.54 & 0.47 & 0.46 & 0.48 & -0.19 & 0.49 & 0.01 & 0.00 & 0.05 & -0.02 & 0.54 & 0.46 & 0.00 & 0.47 & -0.28 & 0.48 \\ 
		9 & 0.2 & 0.1 & 0.05 & 0.25 & 0.49 & 0.04 & 0.51 & -0.10 & 0.69 & 0.01 & 0.00 & 0.01 & -0.15 & 0.02 & 0.00 & -0.01 & 0.01 & -0.02 & 0.01 & 0.00 & -0.01 & 0.01 & -0.17 & 0.01 \\ 
		9 & 0.4 & 0.01 & 0.01 & 0.05 & 0.04 & -0.28 & 0.05 & -0.69 & 0.23 & -0.36 & -0.37 & -0.03 & -0.42 & 0.18 & 0.03 & 0.01 & 0.04 & -0.52 & 0.07 & -0.03 & -0.37 & 0.01 & -0.51 & 0.18 \\ 
		9 & 0.4 & 0.03 & 0.01 & 0.05 & 0.02 & -0.44 & 0.04 & -0.76 & 0.22 & 0.02 & -0.41 & 0.04 & -0.77 & 0.19 & -0.17 & -0.43 & 0.03 & -0.73 & 0.06 & 0.02 & -0.02 & 0.03 & -0.51 & 0.16 \\ 
		9 & 0.4 & 0.1 & 0.01 & 0.05 & 0.02 & -0.38 & 0.04 & -0.84 & 0.22 & 0.02 & 0.01 & 0.03 & -0.75 & 0.20 & -0.37 & -0.40 & 0.03 & -0.74 & 0.06 & -0.04 & -0.38 & 0.01 & -0.75 & 0.18 \\ 
		9 & 0.4 & 0.1 & 0.05 & 0.05 & 0.29 & 0.03 & 0.51 & -0.38 & 0.53 & 0.49 & 0.08 & 0.50 & -0.38 & 0.65 & 0.02 & -0.04 & 0.37 & -0.38 & 0.52 & 0.48 & 0.02 & 0.49 & -0.38 & 0.65 \\ 
		9 & 0.4 & 0.01 & 0.01 & 0.15 & 0.05 & 0.04 & 0.06 & -0.50 & 0.23 & 0.16 & 0.00 & 0.17 & -0.46 & 0.19 & 0.05 & 0.04 & 0.05 & -0.46 & 0.07 & 0.00 & -0.01 & 0.16 & -0.46 & 0.17 \\ 
		9 & 0.4 & 0.03 & 0.01 & 0.15 & 0.05 & -0.27 & 0.05 & -0.63 & 0.23 & 0.17 & 0.16 & 0.18 & -0.31 & 0.19 & 0.03 & -0.40 & 0.05 & -0.73 & 0.06 & 0.15 & 0.01 & 0.16 & -0.47 & 0.17 \\ 
		9 & 0.4 & 0.1 & 0.01 & 0.15 & 0.04 & -0.35 & 0.06 & -0.55 & 0.24 & 0.17 & 0.01 & 0.18 & -0.74 & 0.19 & -0.41 & -0.42 & -0.33 & -0.56 & 0.17 & 0.16 & 0.00 & 0.16 & -0.73 & 0.17 \\ 
		9 & 0.4 & 0.1 & 0.05 & 0.15 & 0.10 & 0.03 & 0.51 & -0.33 & 0.54 & 0.48 & 0.03 & 0.49 & -0.35 & 0.52 & 0.01 & 0.00 & 0.47 & -0.35 & 0.48 & 0.47 & 0.02 & 0.48 & -0.34 & 0.51 \\ 
		9 & 0.4 & 0.01 & 0.01 & 0.25 & 0.07 & 0.06 & 0.17 & -0.31 & 0.21 & 0.01 & -0.29 & 0.16 & -0.46 & 0.17 & -0.15 & -0.31 & 0.04 & -0.47 & 0.24 & 0.01 & 0.00 & 0.15 & -0.47 & 0.16 \\ 
		9 & 0.4 & 0.1 & 0.01 & 0.25 & 0.06 & -0.61 & 0.18 & -0.64 & 0.24 & 0.15 & 0.01 & 0.17 & 0.00 & 0.18 & 0.00 & 0.00 & 0.15 & -0.41 & 0.25 & 0.01 & 0.00 & 0.15 & -0.01 & 0.16 \\ 
		9 & 0.4 & 0.1 & 0.05 & 0.25 & 0.37 & 0.07 & 0.50 & -0.16 & 0.68 & 0.01 & 0.01 & 0.01 & -0.13 & 0.48 & 0.01 & 0.00 & 0.03 & -0.01 & 0.47 & 0.00 & 0.00 & 0.01 & -0.15 & 0.46 \\ 
		10 & 0.02 & 0.1 & 0.05 & 0.05 & 0.14 & 0.03 & 0.45 & -0.27 & 0.66 & 0.44 & 0.02 & 0.50 & -0.35 & 0.63 & 0.44 & 0.02 & 0.46 & -0.35 & 0.64 & 0.44 & 0.03 & 0.49 & -0.36 & 0.64 \\ 
		10 & 0.02 & 0.1 & 0.05 & 0.15 & 0.30 & 0.04 & 0.45 & -0.28 & 0.64 & 0.12 & -0.01 & 0.45 & -0.34 & 0.47 & 0.12 & -0.02 & 0.45 & -0.27 & 0.64 & 0.10 & -0.02 & 0.45 & -0.27 & 0.61 \\ 
		10 & 0.02 & 0.1 & 0.05 & 0.25 & 0.43 & 0.01 & 0.44 & -0.28 & 0.76 & -0.01 & -0.02 & 0.45 & -0.28 & 0.51 & -0.02 & -0.03 & 0.04 & -0.24 & 0.47 & -0.02 & -0.03 & 0.12 & -0.24 & 0.47 \\ 
		10 & 0.2 & 0.1 & 0.01 & 0.05 & -0.07 & -0.47 & 0.01 & -0.87 & 0.04 & -0.02 & -0.51 & -0.01 & -0.95 & 0.01 & -0.13 & -0.46 & -0.01 & -0.93 & 0.03 & -0.02 & -0.03 & -0.01 & -0.51 & 0.00 \\ 
		10 & 0.2 & 0.1 & 0.05 & 0.05 & 0.00 & -0.31 & 0.47 & -0.46 & 0.50 & 0.27 & -0.02 & 0.47 & -0.52 & 0.51 & -0.01 & -0.03 & 0.46 & -0.53 & 0.50 & -0.01 & -0.03 & 0.46 & -0.54 & 0.50 \\ 
		10 & 0.2 & 0.01 & 0.01 & 0.15 & -0.16 & -0.42 & 0.04 & -0.81 & 0.21 & -0.01 & -0.02 & -0.01 & -0.02 & -0.01 & -0.02 & -0.03 & -0.01 & -0.73 & 0.03 & -0.02 & -0.02 & -0.01 & -0.49 & 0.00 \\ 
		10 & 0.2 & 0.03 & 0.01 & 0.15 & -0.15 & -0.69 & -0.02 & -0.82 & 0.03 & -0.02 & -0.02 & -0.01 & -0.82 & 0.00 & -0.02 & -0.46 & -0.01 & -0.79 & 0.05 & -0.02 & -0.02 & -0.01 & -0.81 & -0.00 \\ 
		10 & 0.2 & 0.1 & 0.01 & 0.15 & -0.01 & -0.49 & 0.03 & -0.82 & 0.05 & -0.03 & -0.70 & -0.02 & -0.82 & -0.01 & -0.45 & -0.50 & -0.02 & -0.82 & 0.06 & -0.49 & -0.51 & -0.02 & -0.82 & -0.01 \\ 
		10 & 0.2 & 0.1 & 0.01 & 0.25 & -0.02 & -0.50 & 0.04 & -0.82 & 0.08 & -0.02 & -0.49 & -0.02 & -0.81 & -0.01 & -0.49 & -0.50 & -0.02 & -0.81 & -0.01 & -0.03 & -0.49 & -0.02 & -0.81 & -0.01 \\ 
		10 & 0.2 & 0.1 & 0.05 & 0.25 & -0.16 & -0.24 & -0.02 & -0.35 & 0.47 & -0.02 & -0.02 & -0.01 & -0.33 & 0.52 & -0.02 & -0.02 & -0.02 & -0.34 & 0.02 & -0.02 & -0.02 & -0.02 & -0.35 & 0.02 \\ 
		10 & 0.4 & 0.1 & 0.05 & 0.05 & 0.48 & 0.03 & 0.50 & -0.19 & 0.53 & 0.46 & -0.02 & 0.49 & -0.45 & 0.53 & 0.05 & 0.00 & 0.49 & -0.37 & 0.52 & -0.01 & -0.01 & 0.47 & -0.25 & 0.53 \\ 
		10 & 0.4 & 0.1 & 0.01 & 0.15 & -0.19 & -0.59 & 0.00 & -0.80 & 0.05 & -0.02 & -0.48 & -0.01 & -0.81 & 0.18 & -0.02 & -0.02 & -0.02 & -0.49 & -0.01 & -0.02 & -0.02 & -0.01 & -0.52 & -0.01 \\ 
		\enddata	
	\end{deluxetable*}

	\begin{deluxetable*}{ccccccc|ccccccccccc}
		\centering
		\tablecaption{Distribution of the statistical stellar mass corrections $\Delta(log_{10}(M^*))=log_{10}(M^{*}_{out}/M_{\odot})-log_{10}(M^{*}_{in}/M_{\odot})$ for different output galaxy property combinations and using only 8 NIRCam broad-bands. Columns from 8 to 50 show the normalized distribution. The first row contains the statistical stellar mass corrections intervals considered, from column 8 to 50. A value of -99.9 is present when there are no galaxies for a combination of output galaxy properties. The complete table is available online.\label{tab:PDM_1}}
		\tablecolumns{17}
		\tablenum{4}
		\tablehead{
			\colhead{$<$z$>$} & \colhead{Z} & \colhead{f$_{cov}$}  & \colhead{SFH} & \colhead{t} & \colhead{$<$E(B-V)$>$} & \colhead{S/N$_{F150W}$} & \multicolumn{11}{c}{P($\Delta(log_{10}(M^*))$)} \\
			\colhead{}& \colhead{[Z$_{\odot}$]} & \colhead{} & \colhead{[Gyr]} & \colhead{[Gyr]}  & \colhead{[mag]}  & \colhead{}& \multicolumn{11}{c}{}\\
		}
		\startdata  
		&  &  &  &  &  &  & -1.075 & -1.025 & -0.975 & -0.925 & -0.875 & ... & 0.875 & 0.925 & 0.975 & 1.025 & 1.075 \\
		\hline
		7 & 0.2 & 0 & 0.01 & 0.01 & 0.05 & 10 & 0.00 & 0.00 & 0.00 & 0.00 & 0.00 & ... & 0.00 & 0.00 & 0.00 & 0.00 & 0.00 \\
		7 & 0.2 & 0 & 0.1 & 0.05 & 0.15 & 10 & 0.00 & 0.00 & 0.00 & 0.00 & 0.00 & ... & 0.00 & 0.01 & 0.00 & 0.00 & 0.00 \\
		7 & 0.04 & 1 & 0.01 & 0.01 & 0.25 & 10 & 0.07 & 0.00 & 0.20 & 0.00 & 0.67 & ... & 0.00 & 0.00 & 0.00 & 0.00 & 0.00 \\
		\enddata	
	\end{deluxetable*}
	
	\begin{deluxetable*}{ccccccc|ccccccccccc}
		\centering
		\tablecaption{Distribution of the statistical stellar mass corrections $\Delta(log_{10}(M^*))=log_{10}(M^{*}_{out}/M_{\odot})-log_{10}(M^{*}_{in}/M_{\odot})$ for different output galaxy property combinations and using only 8 NIRCam broad-bands and MIRI/F560W. Columns from 8 to 50 show the normalized distribution. The first row contains the statistical stellar mass correction intervals considered, from column 8 to 50. A value of -99.9 is present when there are no galaxies for a combination of output galaxy properties. The complete table is available online.\label{tab:PDM_2}}
		\tablecolumns{17}
		\tablenum{5}
		\tablehead{
			\colhead{$<$z$>$} & \colhead{Z} & \colhead{f$_{cov}$}  & \colhead{SFH} & \colhead{t} & \colhead{$<$E(B-V)$>$} & \colhead{S/N$_{F150W}$} & \multicolumn{11}{c}{P($\Delta(log_{10}(M^*))$)} \\
			\colhead{}& \colhead{[Z$_{\odot}$]} & \colhead{} & \colhead{[Gyr]} & \colhead{[Gyr]}  & \colhead{[mag]}  & \colhead{}& \multicolumn{11}{c}{}\\
		}
		\startdata  
		&  &  &  &  &  & & -1.075 & -1.025 & -0.975 & -0.925 & -0.875 & ... & 0.875 & 0.925 & 0.975 & 1.025 & 1.075 \\
		\hline
		7 & 0.2 & 0 & 0.01 & 0.01 & 0.05 & 10 & 0.00 & 0.00 & 0.00 & 0.00 & 0.00 & ... & 0.00 & 0.00 & 0.00 & 0.00 & 0.00 \\
		7 & 0.2 & 0 & 0.1 & 0.01 & 0.15 & 10 & 0.00 & 0.00 & 0.00 & 0.00 & 0.01 & ... & 0.00 & 0.00 & 0.00 & 0.00 & 0.00 \\
		7 & 0.2 & 0 & 0.01 & 0.01 & 0.25 & 10 & 0.01 & 0.00 & 0.00 & 0.00 & 0.00 & ... & 0.00 & 0.00 & 0.00 & 0.00 & 0.00 \\
		\enddata	
	\end{deluxetable*}
	
	\begin{deluxetable*}{ccccccc|ccccccccccc}
		\centering
		\tablecaption{Distribution of the statistical stellar mass corrections $\Delta(log_{10}(M^*))=log_{10}(M^{*}_{out}/M_{\odot})-log_{10}(M^{*}_{in}/M_{\odot})$ for different output galaxy property combinations and using only 8 NIRCam broad-bands and MIRI/F770W. Columns from 8 to 50 show the normalized distribution. The first row contains the statistical stellar mass correction intervals considered, from column 8 to 50. A value of -99.9 is present when there are no galaxies for a combination of output galaxy properties. The complete table is available online.\label{tab:PDM_3}}
		\tablecolumns{17}
		\tablenum{6}
		\tablehead{
			\colhead{$<$z$>$} & \colhead{Z} & \colhead{f$_{cov}$} & \colhead{SFH} & \colhead{t} & \colhead{$<$E(B-V)$>$} & \colhead{S/N$_{F150W}$} & \multicolumn{11}{c}{P($\Delta(log_{10}(M^*))$)} \\
			\colhead{}& \colhead{[Z$_{\odot}$]} & \colhead{} & \colhead{[Gyr]} & \colhead{[Gyr]}  & \colhead{[mag]}  & \colhead{}& \multicolumn{11}{c}{}\\
		}
		\startdata  
		&  &  &  &  &  &  &-1.075 & -1.025 & -0.975 & -0.925 & -0.875 & ... & 0.875 & 0.925 & 0.975 & 1.025 & 1.075 \\
		\hline
		7 & 0.2 & 0 & 0.01 & 0.01 & 0.05 & 10 & 0.00 & 0.00 & 0.00 & 0.00 & 0.00 & ... & 0.00 & 0.00 & 0.00 & 0.00 & 0.00 \\
		7 & 0.2 & 1 & 0.1 & 0.05 & 0.05 & 10 & 0.00 & 0.00 & 0.00 & 0.00 & 0.00 & ... & 0.00 & 0.01 & 0.01 & 0.00 & 0.00 \\
		7 & 0.2 & 0 & 0.01 & 0.01 & 0.25 & 10 & 0.01 & 0.00 & 0.01 & 0.00 & 0.00 & ... & 0.00 & 0.00 & 0.00 & 0.00 & 0.00 \\
		\enddata	
	\end{deluxetable*}
	
	\begin{deluxetable*}{ccccccc|ccccccccccc}
		\centering
		\tablecaption{Distribution of the stellar mass errors $\Delta(log_{10}(M^*))=log_{10}(M^{*}_{out}/M_{\odot})-log_{10}(M^{*}_{in}/M_{\odot})$ for different output galaxy property combinations and using only 8 NIRCam broad-bands, MIRI/F560W and MIRI/F770W. Columns from 8 to 50 show the normalized distribution. The first row contains the stellar mass error intervals considered, from column 8 to 50. A value of -99.9 is present when there are no galaxies for a combination of output galaxy properties. The complete table is available online.\label{tab:PDM_4}}
		\tablecolumns{17}
		\tablenum{7}
		\tablehead{
			\colhead{$<$z$>$} & \colhead{Z} & \colhead{f$_{cov}$} & \colhead{SFH} & \colhead{t} & \colhead{$<$E(B-V)$>$} & \colhead{S/N$_{F150W}$}  & \multicolumn{11}{c}{P($\Delta(log_{10}(M^*))$)} \\
			\colhead{}& \colhead{[Z$_{\odot}$]} & \colhead{} & \colhead{[Gyr]} & \colhead{[Gyr]}  & \colhead{[mag]}   & \colhead{}& \multicolumn{11}{c}{}\\
		}
		\startdata  
		&  &  &  &  &  &  & -1.075 & -1.025 & -0.975 & -0.925 & -0.875 & ... & 0.875 & 0.925 & 0.975 & 1.025 & 1.075 \\
		\hline
		7 & 0.2 & 0 & 0.01 & 0.01 & 0.05 & 10 & 0.00 & 0.00 & 0.00 & 0.00 & 0.00 & ... & 0.00 & 0.00 & 0.00 & 0.00 & 0.00 \\
		7 & 0.2 & 1 & 0.1 & 0.05 & 0.05 & 10  & 0.00 & 0.00 & 0.00 & 0.00 & 0.00 & ... & 0.00 & 0.01 & 0.01 & 0.00 & 0.00 \\
		7 & 0.2 & 0 & 0.01 & 0.01 & 0.25 & 10 & 0.01 & 0.00 & 0.01 & 0.00 & 0.00 & ... & 0.00 & 0.00 & 0.00 & 0.00 & 0.00 \\
		\enddata	
	\end{deluxetable*}
	
	\section{Comparison with the JAGUAR Catalogue}\label{sec:comparison}
	We test our method by applying the statistical stellar mass correction to the NIRCam mock observations derived as part of the JWST Advanced Deep Extragalactic Survey (JADES) using the JAdes extraGalactic Ultradeep Artificial Realizations (JAGUAR) package \citep{Williams2018}. These mock catalogues are derived by generating a galaxy population that follows empirical functions, such as the observed stellar mass and luminosity functions at different redshifts, and assigning to each of them a SED template using the BEAGLE tool \citep[for BayEsian Analysis of GaLaxy sEds;][]{Chevallard2016}. We consider only mock galaxies between z$=$6.5 and 10.5 for a total of 23493 objects. We consider an observational depth of 10 $\sigma$ at 29 AB mag at 1.5 $\mu$m and we consider only galaxies with S/N$_{F150W}\geq$10.\par
\begin{figure}[ht!]
	\center{
		\includegraphics[trim={0cm 5cm 0cm 5cm},clip,width=1\linewidth, keepaspectratio]{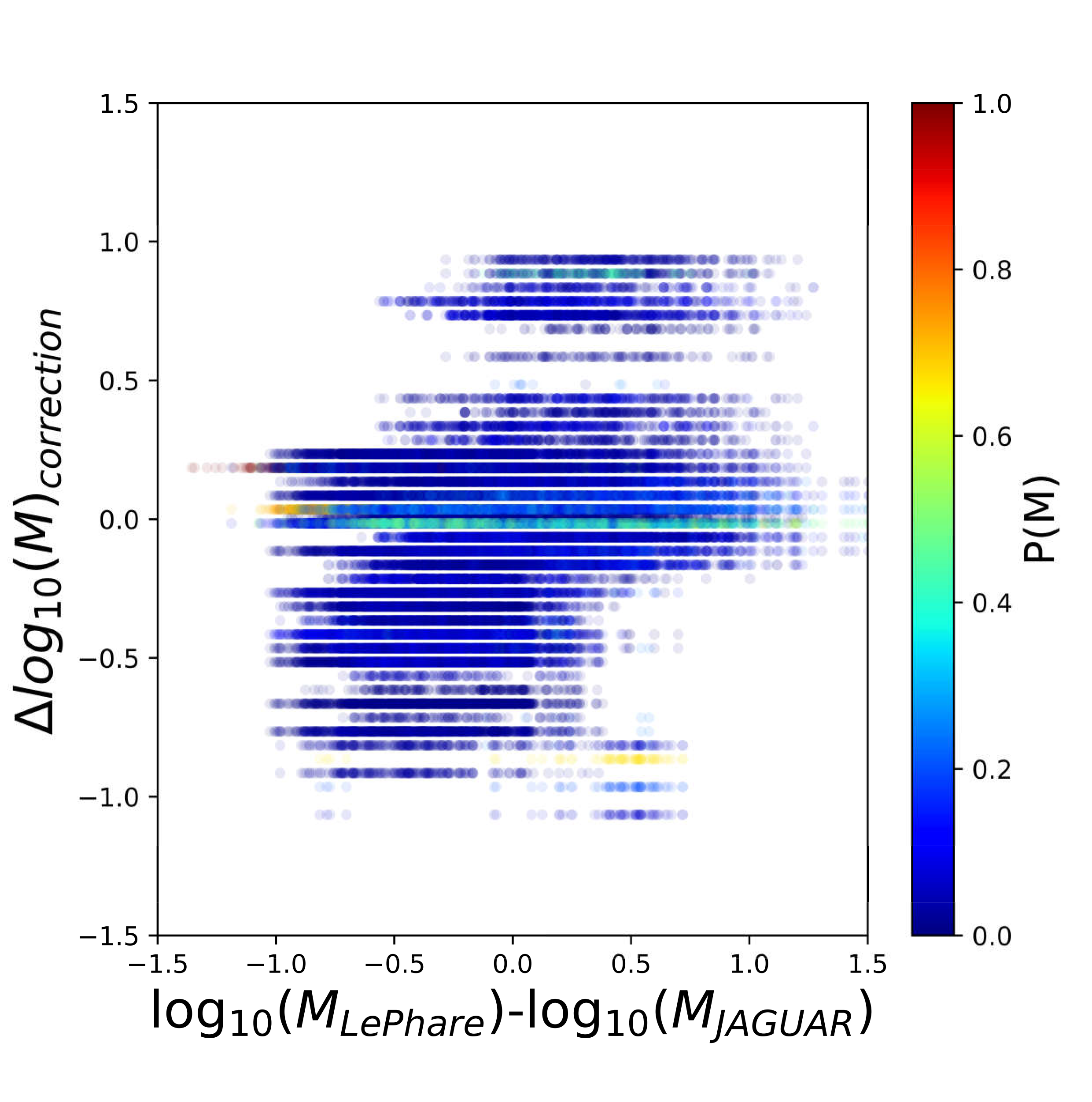}
	\caption{Comparison between the stellar mass correction and the difference between the JAGUAR stellar mass and the stellar mass recovered with \textit{LePhare}. We color-coded each correction by its probability (Table \ref{tab:PDM_1}).\label{fig:DM-JAGUAR_comparison}}}
	\end{figure}

 
	We run the code \textit{LePhare} using the same templates described in section \ref{sec:PropDerivation}. We then correct the stellar mass function in the different redshift bins, by considering the full stellar mass offset distribution associate with the best template obtained for each object (Table \ref{tab:PDM_1}). In Figure \ref{fig:DM-JAGUAR_comparison}, we compare the expected stellar mass corrections and the difference between the original and the recovered stellar mass of the JAGUAR catalogue, before applying any correction. The overall magnitudes of the corrections are similar to the stellar mass discrepancies and, even with a large scatter, they show a positive correlation, showing that the stellar mass correction generally improves the stellar mass estimation obtained with \textit{LePhare} and the templates considered in this work. In line with the detail analysis of the main degeneracies done in Sec. \ref{sec:results}, we notice that solar metallicity systems tend to have a correct stellar mass estimation, while the stellar mass estimation is more challenging for objects with high star-formation-rates that present, indeed, emission lines and objects with t$_{out}<$0.2 Gyr. \par
\begin{figure*}[ht!]
		\center{
			\includegraphics[trim={0cm 0cm 0cm 0cm},clip,width=0.48\linewidth, keepaspectratio]{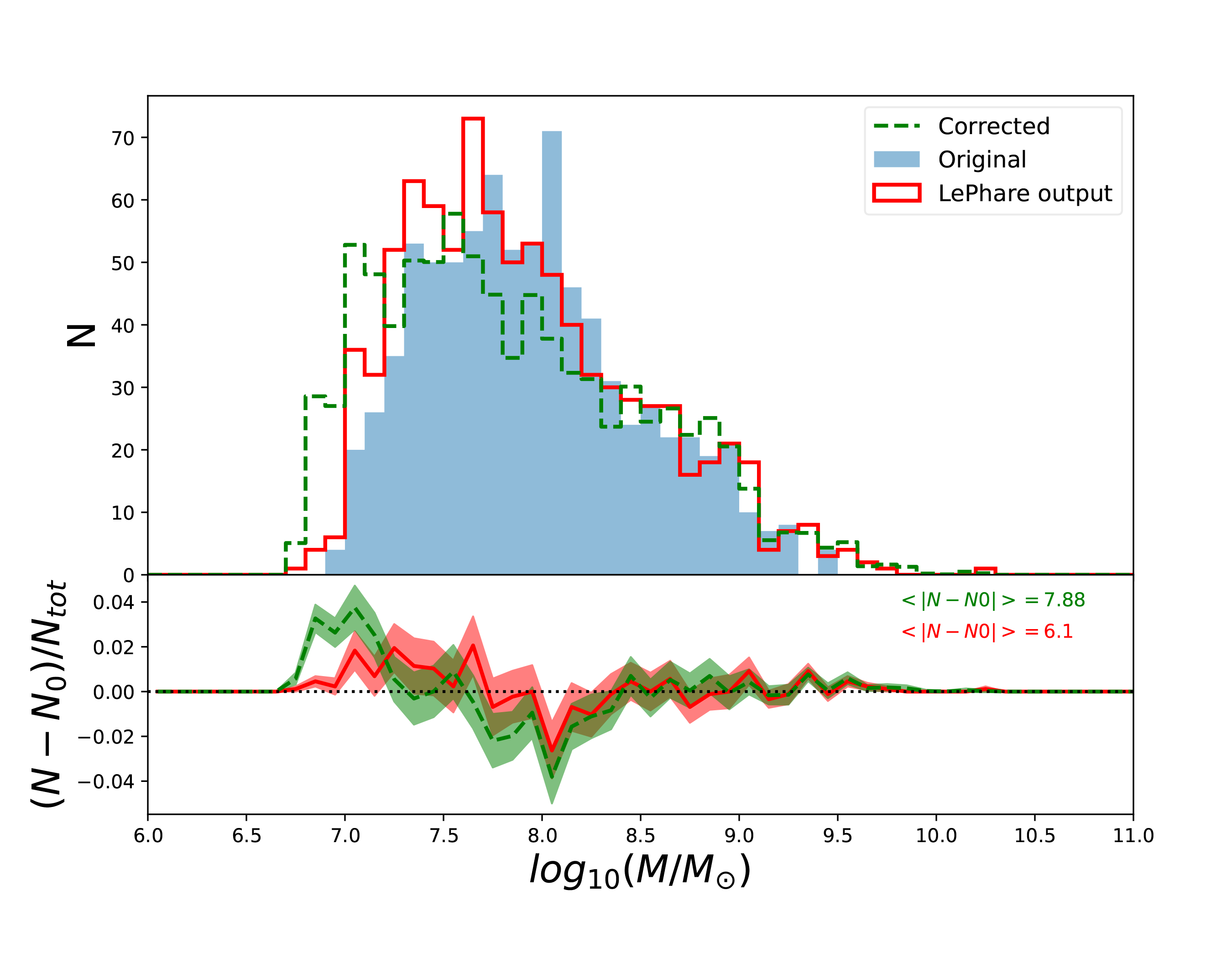}
			\includegraphics[trim={0cm 0cm 0cm 0cm},clip,width=0.48\linewidth, keepaspectratio]{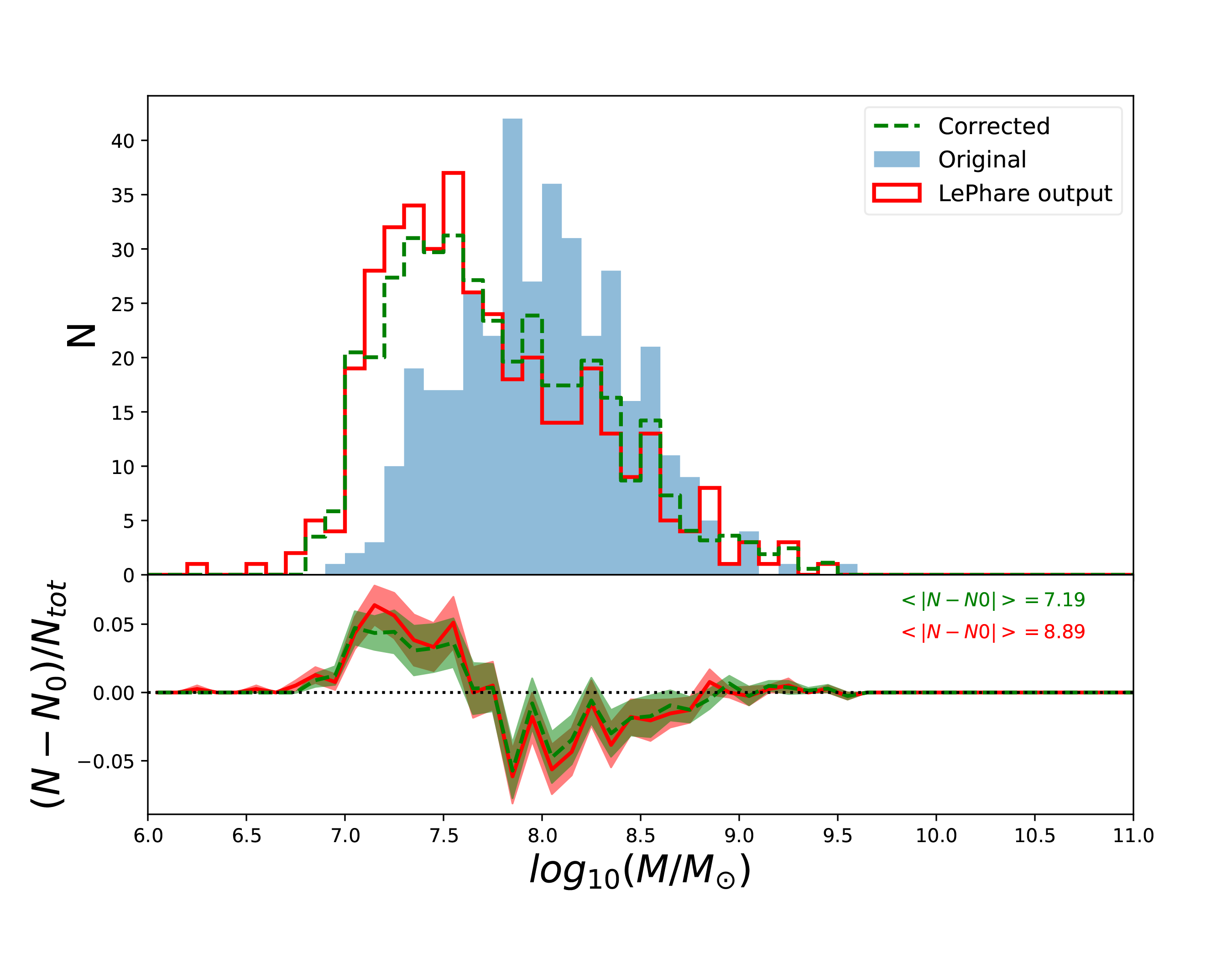}
			\includegraphics[trim={0cm 0cm 0cm 0cm},clip,width=0.48\linewidth, keepaspectratio]{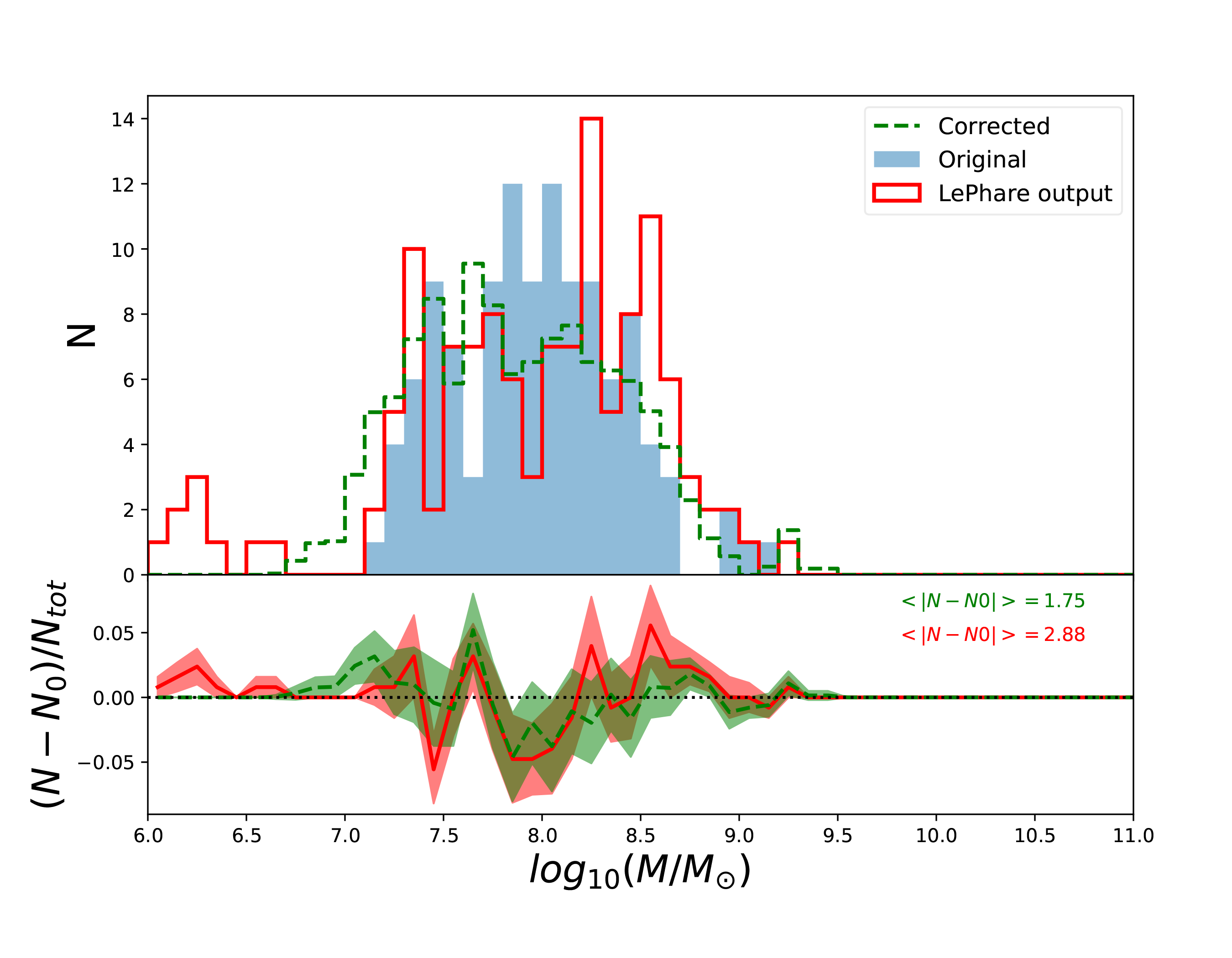}
			\includegraphics[trim={0cm 0cm 0cm 0cm},clip,width=0.48\linewidth, keepaspectratio]{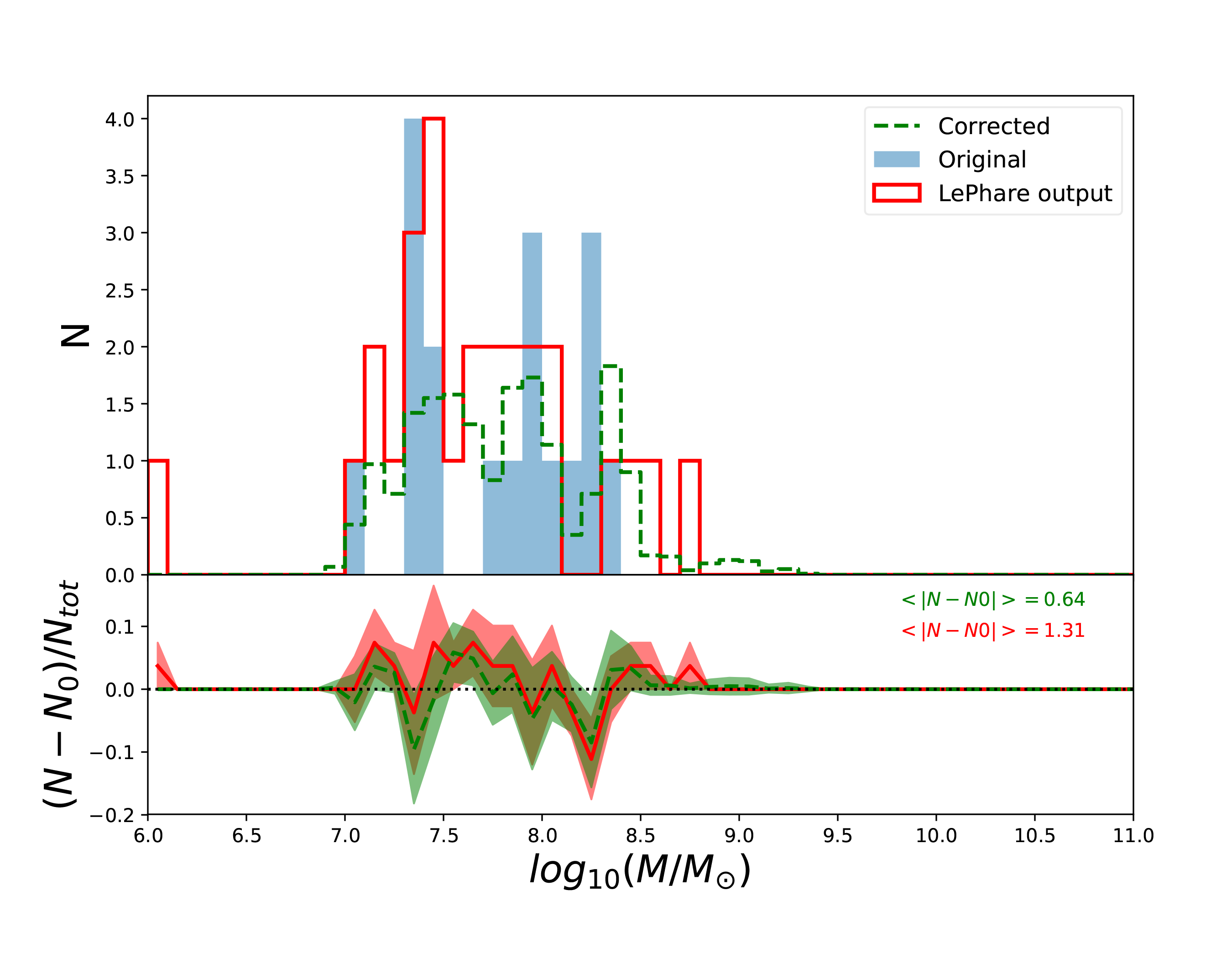}}
		\caption{\textit{Upper panels: }Stellar mass distribution in the JAGUAR catalogue (\textit{Filled blue}) and stellar mass distribution derived with the \textit{Yggddrasil} templates before (\textit{red line}) and after the statistical stellar mass correction (\textit{dashed green line}). \textit{Bottom panels: } offsets between the original JAGUAR stellar mass and the stellar mass derived with the \textit{Yggddrasil} templates before (\textit{red line}) and after the statistical stellar mass correction (\textit{dashed green line}) and the average values of the absolute number differences are shown on the right. Shadowed coloured areas shows the statistical uncertainties. Both stellar mass distributions  and offsets are shown for four redshift bins: 6.5$\leq z <$7.5 (\textit{upper left}), 7.5$\leq z <$8.5 (\textit{upper right}), 8.5$\leq z <$9.5 (\textit{bottom left}) and 9.5$\leq z <$10.5 (\textit{bottom right}). \label{fig:DM-JAGUAR}}
	\end{figure*}

 
	Figure \ref{fig:DM-JAGUAR} shows the difference between the stellar mass distribution of the JAGUAR catalogue, the one derived using \textit{LePhare} and the Yggdrasil templates considered in this work before and after applying the statistical errors. Excluding z$=$7, the mean absolute error on the stellar mass distributions decreases by 20-50$\%$ after applying the stellar mass corrections derived here. It is evident that, for this specific case, the main improvement on applying the statistical stellar mass correction consist on not overestimating the number of extremely low-stellar mass objects (M$^*<10^{7} M_{\odot}$) that are indeed expected to be an important fraction of the galaxy population that will be observed with \textit{JWST} at z$>$7 \citep{Ceverino2019}.  This may be an important issue for future works that aim on recovering the proper value of the stellar mass function faint-end slope at high-z. \par
	As already mentioned, the corrections derived in this work are not exhaustive. Indeed, an additional offset between the original stellar mass and the one derived with \textit{Yggdrasil} templates is present and it is due to different emission-line equivalent width recipes and SFH used in \textit{Yggdrasil} and BEAGLE.\par
	Overall, the statistical stellar mass corrections improve the stellar mass distribution derivation obtained with the \textit{Yggdrasil} templates, even if the galaxies have completely different SFH and emission-line recipes. 
	
	\section{Conclusions}\label{sec:conclusions}
	In a series of papers, of which this is the third, we have created and analyzed a mock galaxy sample to investigate how different galaxy properties will be derived using the \textit{JWST} NIRCam broad-band filters and the two MIRI filters F560W and F770W for galaxies at redshifts between 7 and 10. In particular, we consider a mock galaxy sample with good photometry (S/N$_{F150W}>$10) for which the redshift is well recovered and we derive galaxy properties fitting the broad-band SED, in a similar way to what will be done for galaxies in future \textit{JWST} deep blank-field imaging surveys. Then, we compare the derived galaxy properties with the input ones to understand biases in the galaxy property estimation. \par
	
	In this paper, we focus on the stellar mass estimation, providing a list of statistical stellar mass corrections to take into account when estimating stellar masses of large samples of high-z galaxies that will be observed in the next future with \textit{JWST}. These corrections can be used directly when stellar masses are derived using the \textit{LePhare} code and the SED template considered here and they can be generally considered as an indication of the possible template degeneracies affecting the stellar mass derivation.
	In particular, we provide the 25$\%$, 50$\%$ and 75$\%$ quartiles, minimum and maximum value of the stellar mass offsets, together with the full stellar mass offset distributions, for each combination of output galaxy parameter and for different \textit{JWST} broad-band filter combinations. \par
	Median stellar mass offsets are generally small, but the 25$\%$ and 75$\%$ quartiles for some specific templates range from -0.83 dex to 0.88 dex, therefore some galaxy models may need significant stellar mass corrections.
	\par

	In particular, we notice that:
	\begin{itemize}
		\item galaxies without nebular emission lines, in output, that were originally with emission lines have overestimated SFH and age. This results in an overestimation of the stellar mass that is even 10 times the original stellar mass. On the other hand, galaxies with nebular emission lines, in output, that were originally without emission lines have stellar masses that are generally underestimated, even down to 10 times the original mass.
		
		\item the stellar mass of galaxies is more difficult to estimate at z$=$10, when only 8 NIRCam bands are considered, but also at z$=$7. The first effect is due to the fact that no NIRCam bands purely cover $\lambda>4000\,\AA$ break, therefore MIRI bands are necessary to trace these wavelengths and improve the stellar mass estimation. The reduced age-extinction degeneracy is instead one of the reasons why the stellar mass recovery is less difficult at z$>$7 than at z$=$7.
		
		\item Stellar masses are well recovered for galaxies with output color excess between 0.2 and 0.3 mag, i.e. the maximum value analyzed in this paper, while the stellar mass estimation is less accurate for relative dust-free templates.
		
		\item For galaxies without emission lines, i.e. quenched or star-forming galaxies with f$_{cov}=$0, the stellar mass is generally well recovered, except for the lowest metallicity considered, which is 1/50 Z$_{\odot}$.
		
		\item we apply the statistical stellar mass corrections presented in this work to the JAGUAR catalogue, which has been derived using different assumptions \citep{Williams2018}. The discrepancies in the recovered stellar mass distribution decrease by 20-50$\%$ at z$>$7 when including the stellar mass corrections. Results highlight the importance of the considered statistical stellar mass corrections to properly derive the distribution of the low-mass galaxy population that otherwise tends to be overestimated in number. This is indeed essential for future works that aim at deriving the faint-end slope of the high-z stellar mass function.
		
	\end{itemize}
	Overall, the stellar mass estimation is challenging for young galaxies with nebular emission lines or galaxies with emission lines that have not been properly identified.  Adding at least one of the two MIRI bands at shortest wavelengths improve the stellar mass estimation, refining the average estimation or reducing the worst outliers. \par 
 In the future, additional investigations are necessary to expand this analysis on extremely dusty objects and galaxies with a not-negligible contribution by active galactic nuclei.

\vspace{5mm}	
	\acknowledgments
	LB and KIC acknowledge funding from the European Research Council through the award of the Consolidator Grant ID 681627-BUILDUP. PGP-PG acknowledges support from the Spanish Government grant AYA2015-63650-P. LC acknowledges support by the Spanish Ministry for Science, Innovation and Universities under project ESP2017-83197.
	
\software{Yggdrasill (Zackrisson et al. 2011), LePhare (Arnouts et al. 1999; Ilbert et al. 2006)}

	\vspace{5mm}
	
	
	


\begin{thebibliography}{}
		%
		\bibitem[Arnouts et al.(1999)]{Arnouts1999} Arnouts, S., Cristiani, S., Moscardini, L., et al.\ 1999, \mnras, 310, 540
		\bibitem[Bisigello et al.(2016)]{Bisigello2016} Bisigello, L., Caputi, K.~I., Colina, L., et al.\ 2016, \apjs, 227, 19 
		\bibitem[Bisigello et al.(2017)]{Bisigello2017} Bisigello, L., Caputi, K.~I., Colina, L., et al.\ 2017, \apjs, 231, 3 
		\bibitem[Bisigello et al.(2018)]{Bisigello2018} Bisigello, L., Caputi, K.~I., Grogin, N., \& Koekemoer, A.\ 2018, \aap, 609, A82 
		\bibitem[Boogaard et al.(2018)]{Boogaard2018} Boogaard, L.~A., Brinchmann, J., Bouch{\'e}, N., et al.\ 2018, arXiv:1808.04900 
		\bibitem[Brinchmann et al.(2004)]{Brinchmann2004} Brinchmann, J., Charlot, S., White, S.~D.~M., et al.\ 2004, \mnras, 351, 1151
		\bibitem[Bruzual \& Charlot(2003)]{B&C2003} Bruzual, G., \& Charlot, S.\ 2003, \mnras, 344, 1000 
		\bibitem[Calzetti et al.(2000)]{Calzetti2000} Calzetti, D., Armus, L., Bohlin, R.~C., et al.\ 2000, \apj, 533, 682 
		\bibitem[Caputi et al.(2015)]{Caputi2015} Caputi, K.~I., Ilbert, O., Laigle, C., et al.\ 2015, \apj, 810, 73 
		\bibitem[Caputi et al.(2017)]{Caputi2017} Caputi, K.~I., Deshmukh, S., Ashby, M.~L.~N., et al.\ 2017, \apj, 849, 45
		\bibitem[Ceverino et al. (2019)]{Ceverino2019} Ceverino, D., Klessen, R.~S., \& Glover, S.~C.~O.\ 2019, \mnras, 484, 1366 
		\bibitem[Chevallard \& Charlot(2016)]{Chevallard2016} Chevallard, J., \& Charlot, S.\ 2016, \mnras, 462, 1415  
		\bibitem[Erb et al.(2006)]{Erb2006} Erb, D.~K., Shapley, A.~E., Pettini, M., et al.\ 2006, \apj, 644, 813
		\bibitem[Gardner et al.(2009)]{Gardner2009} Gardner, J.~P., Mather, J.~C., Clampin, M., et al.\ 2009, \assp, 10, 1  
		\bibitem[Ilbert et al.(2006)]{Ilbert2006} Ilbert, O., Arnouts, S., McCracken, H.~J., et al.\ 2006, \aap, 457, 841
		\bibitem[Kroupa(2002)]{Kroupa2002} Kroupa, P.\ 2002, Science, 295, 82 
		\bibitem[Maier et al.(2015)]{Maier2015} Maier, C., Ziegler, B.~L., Lilly, S.~J., et al.\ 2015, \aap, 577, A14 
		\bibitem[Maiolino et al.(2008)]{Maiolino2008} Maiolino, R., Nagao, T., Grazian, A., et al.\ 2008, \aap, 488, 463
		\bibitem[Noeske et al.(2007)]{Noeske2007} Noeske, K.~G., Weiner, B.~J., Faber, S.~M., et al.\ 2007, \apjl, 660, L43 
		\bibitem[Oke \& Gunn(1983)]{Oke1983} Oke, J.~B., Gunn, J.~E., 1983, \apj, 266, 713
		\bibitem[Peng et al.(2010)]{Peng2010} Peng, Y.-j., Lilly, S.~J., Kova{\v c}, K., et al.\ 2010, \apj, 721, 193 
		\bibitem[Rieke et al.(2005)]{Rieke2005} Rieke, M.~J., Kelly, D. \& Horner, S. \ 2005, SPIE, 5904, 1 
		\bibitem[Rieke et al.(2015)]{Rieke2015} Rieke, G.~H., Wright, G.~S., B\"oker, T., et al. \ 2015,  \pasp, 127, 584
		\bibitem[Rodighiero et al.(2011)]{Rodighiero2011} Rodighiero, G., Daddi, E., Baronchelli, I., et al.\ 2011, \apjl, 739, L40 
		\bibitem[Santini et al.(2015)]{Santini2015} Santini, P., Ferguson, H.~C., Fontana, A., et al.\ 2015, \apj, 801, 97 
		\bibitem[Stark et al.(2013)]{Stark2013} Stark, D.~P., Schenker, M.~A., Ellis, R., et al.\ 2013, \apj, 763, 129 
		\bibitem[Tasca et al.(2015)]{Tasca2015} Tasca, L.~A.~M., Le F{\`e}vre, O., Hathi, N.~P., et al.\ 2015, \aap, 581, A54 
		\bibitem[Tremonti et al.(2004)]{Tremonti2004} Tremonti, C.~A., Heckman, T.~M., Kauffmann, G., et al.\ 2004, \apj, 613, 898 
		\bibitem[Williams et al.(2018)]{Williams2018} Williams, C.~C., Curtis-Lake, E., Hainline, K.~N., et al.\ 2018, \apjs, 236, 33 
		\bibitem[Whitaker et al.(2014)]{Whitaker2014} Whitaker, K.~E., Franx, M., Leja, J., et al.\ 2014, \apj, 795, 104 
		\bibitem[Wright et al.(2015)]{Wright2015} Wright, G.~S., Wright, D., Goodson, G.~B., et al.\ 2015, \pasp, 127, 595 
		\bibitem[Zackrisson et al.(2011)]{Zackrisson2011} Zackrisson, E., Rydberg, C.-E., Schaerer, D., {\"O}stlin, G., \& Tuli, M.\ 2011, \apj, 740, 13
		
		
	\end{thebibliography}
\end{document}